%% file: main.tex
\documentclass[a4 paper, 11pt]{article}
\pdfoutput=1
\usepackage{comment}
\usepackage{jheppub}
\usepackage{subcaption}
\usepackage{amsmath}
\usepackage{mathrsfs}
\usepackage{amssymb}
\usepackage{amscd}
\usepackage{dsfont}
\usepackage{enumerate}
\usepackage{amsfonts}
\usepackage{epsfig}
\usepackage{mathtools}
\usepackage{yfonts}
\usepackage{bbold}
\usepackage{breqn}
\usepackage[utf8]{inputenc}
\usepackage[english]{babel}
\usepackage{graphicx}
\usepackage{slashed}
\allowdisplaybreaks[4]

\input{preamble.tex}

\title{Double-Current Deformations of Two-Dimensional QFTs with Anomalies}

\author{Zhengyuan Du$^{\gamma,\tau}$}

\emailAdd{dzy.2021@tsinghua.org.cn}

\abstract
{
We construct the double-current deformations of two-dimensional quantum field
theories whose partition functions have background gauge-field
anomalies.  Extending the path integral construction of
\cite{Dubovsky:2023lza}, we couple the seed theory to dynamical gauge fields and
compact Stueckelberg fields and insert parallel transport in the anomaly line bundle.
The deformed partition function then has the same anomaly as the undeformed one.

For flat background gauge fields, the Stueckelberg non-zero modes localize the
dynamical gauge fields to flat connections, reducing the deformation to a
finite-dimensional holonomy integral.  We derive the integral kernel on the torus and its
higher-genus generalization.  For the compact boson, or
equivalently the Abelian $U(1)$ WZW model, the kernel gives a Gaussian
transform of the torus partition function: at zero background the spectrum is
obtained by $k\to K_\lambda$, while contact terms and spectral-flow data remain
controlled by the original anomaly.  As anomaly-free massive examples, we apply the
kernel to massive complex bosons and massive Dirac fermions, for which the
finite-volume spectra are obtained from the undeformed twisted spectra by
charge-dependent shifts of the twists.  We also formulate the anomaly-compatible non-Abelian and homogeneous
Yang-Baxter generalization.
}

\begin{document}
\maketitle
\section{Introduction}
\label{sec:introduction}

Solvable deformations provide a controlled way of moving away from a seed
two-dimensional QFT while retaining exact information about the deformed theory.
The prototype is the \(T\bar{T}\) deformation
\cite{Zamolodchikov:2004ce,Smirnov:2016lqw,Cavaglia:2016oda}, see \cite{Jiang:2019epa} for a review.  Although this is
an irrelevant deformation, its finite-volume spectrum \cite{Smirnov:2016lqw,Cavaglia:2016oda}, scattering amplitudes \cite{Dubovsky:2012wk,Dubovsky:2013ira,Dubovsky:2017cnj} and
torus partition functions \cite{Datta:2018thy,Aharony:2018bad} are governed by universal formulae, and its deformed conformal symmetries \cite{Guica:2022gts} and operators can be organized in
non-local variables \cite{Aharony:2023dod,Chen:2025jzb}.  A natural way to understand this
structure is through the path integral.  In the \(T\bar{T}\) case the
deformation can be formulated by coupling the seed theory to a topological
gravity sector \cite{Dubovsky:2017cnj,Dubovsky:2018bmo,Cardy:2018sdv}.  The auxiliary fields play the role of dynamical fields, and the finite
deformation is therefore defined by a gauging procedure rather than only by a
formal composite-operator perturbation.

The same point of view applies to deformations generated by conserved currents.
The \(J\bar T\) and \(T\bar J\) deformations are Lorentz-breaking but remain
solvable examples \cite{Guica:2017lia}.  Their finite-volume spectra
\cite{Guica:2017lia,Chakraborty:2018vja}, modular properties
\cite{Aharony:2018ics}, deformed symmetries \cite{Georgescu:2024ppd}, and
operators written in non-local variables \cite{Guica:2021fkv,Chen:2025umo}
are under good control, and they also admit topological-gravity realizations
\cite{Anous:2019osb,Aguilera-Damia:2019tpe}.
When the seed theory is conformal, current-current deformations include the
familiar marginal \(J\bar{J}\)  \cite{Chaudhuri:1988qb,Cardy:2019qao,Borsato:2023dis}, whose torus and higher-genus partition functions admit kernel-integral descriptions \cite{Liu:2026zbp}.  In holographic examples based on
strings on \(AdS_3\), TsT transformations \cite{Lunin:2005jy} of the bulk background are dual to the
single-trace \(T\bar{T}\)-like deformations of the boundary CFT, including
\(T\bar{T}\), \(J\bar{T}\), \(T\bar{J}\), and $J\bar{J}$
deformations in sectors
\cite{Giveon:2017nie,Kutasov:2017myj,Kutasov:2019mdf,Apolo:2018qpq,Apolo:2019zai,Apolo:2021wcn}.
Beyond the Abelian TsT constructions, homogeneous Yang-Baxter deformations provide a natural non-Abelian extension. In the double-current language, this extension is realized through Deformed T-dual Models associated with symmetry algebras admitting non-trivial central extensions, which were constructed in \cite{Borsato:2016pas,Borsato:2017qsx} and shown to be equivalent to homogeneous Yang-Baxter deformations \cite{Klimcik:2002zj,Klimcik:2008eq,Borsato:2018spz,Borsato:2023dis,Delduc:2013qra,Kawaguchi:2014qwa}.

The field-theoretic construction of \cite{Dubovsky:2023lza} gives a unified
path integral definition of Abelian double-current deformations and explains the
relation of the non-Abelian extension to homogeneous Yang-Baxter deformations.
Suppose, for the moment, that the seed theory has two commuting \(U(1)\) global
symmetries, with conserved currents \(J^a\), \(a=1,2\).  The infinitesimal
deformation is generated by
\begin{equation}
    \frac{\partial S_\lambda}{\partial\lambda}
    \sim
    i\int_\Sigma \epsilon_{ab} J_\lambda^a\wedge J_\lambda^b .
    \label{eq:intro-current-current-flow}
\end{equation}
The finite definition is a path integral over auxiliary gauge fields for these
two symmetries. The important
point for us is that this definition is more general than the conformal or
sigma-model examples: it can be stated for any two-dimensional QFT with the
relevant global symmetries, without assuming integrability, conformal
invariance, or a Lagrangian sigma-model description.  This is the starting point
of the present work.

When the two global symmetries of the seed theory are coupled to background
gauge fields, the corresponding partition function need not be gauge invariant.
Under a finite background gauge transformation \(U\), we allow
\begin{equation}
    Z_0[A^U]
    =
    \exp\!\left[-\mathcal I_{\rm anom}[U,A]\right]Z_0[A],
    \label{eq:intro-finite-anomaly}
\end{equation}
where \(\mathcal I_{\rm anom}\) obeys the Wess-Zumino consistency condition
\cite{Wess:1971yu,Bismut:1986wr,Freed:1992vw}.  Equivalently, the anomalous
partition function is not an ordinary function on the space of background
gauge fields, but a section of an anomaly line bundle.  If one repeats the
anomaly-free gauging construction naively, the integrand does not descend to
the auxiliary gauge quotient.  Our prescription is instead to perform the
deformation with this anomaly line bundle held fixed.

Let \(B\) be the dynamical gauge field, let \(h\) denote compact Stueckelberg
fields, and let \(B_h\) be the corresponding auxiliary-gauge-invariant
combination.  We set
\[
    \Theta^a = A^a - B_h^a
\]
and define, schematically,
\begin{equation}
\begin{aligned}
    Z_\lambda[A]
    =
    \int\frac{[D\Phi\,DB\,Dh]}{\mathrm{Vol}(\mathcal G)}
    \,
    T[A,B_h]\,
    \exp\!\left[
    -S_0[\Phi|B]
    -\mathcal I_{\rm anom}[h,B]
    -\frac{i}{4\lambda}\int_\Sigma\epsilon_{ab}\Theta^a\wedge\Theta^b
    \right].
\end{aligned}
    \label{eq:intro-anomalous-transform}
\end{equation}
The last term is the anomaly-free double-current kernel.  The new ingredients
are the Stueckelberg anomaly factor and the transport factor \(T[A,B_h]\),
which implements parallel transport in the anomaly line bundle from the fiber
over \(B_h\) to the fiber over \(A\).  Its endpoint transformation law is fixed
by the finite anomaly,
\begin{equation}
    T[A^V,B^V]
    =
    \exp\!\left[
    -\mathcal I_{\rm anom}[V,A]
    +
    \mathcal I_{\rm anom}[V,B]
    \right]T[A,B],
    \qquad
    T[A,A]=1 .
    \label{eq:intro-transport-endpoint-law}
\end{equation}
Together with the Stueckelberg compensator, this endpoint law ensures that the
deformed partition function transforms with exactly the same anomaly as the
seed theory,
\begin{equation}
    Z_\lambda[A^V]
    =
    \exp\!\left[-\mathcal I_{\rm anom}[V,A]\right]Z_\lambda[A].
    \label{eq:intro-deformed-anomaly}
\end{equation}
The anomaly data are therefore held fixed along the deformation.  Since the
Stueckelberg compensator and the transport factor are independent of
\(\lambda\), the \(\lambda\)-dependence of \(Z_\lambda[A]\) comes entirely from
the quadratic topological kernel.  Thus the deformation still admits a
double-current interpretation.  In the anomalous theory, however, the current
defined by the kernel need not coincide with the physical background-field
current; their difference is controlled by the endpoint response of the
anomaly-line-bundle transport.  We return to this distinction in
Section~\ref{subsec:current-current-interpretation}.

For flat background gauge fields, the path integral reduces to a
finite-dimensional transform over holonomies.  On the torus, the \(A\)-cycle
holonomies specify twists, while the \(B\)-cycle holonomies play the role of
chemical potentials. The non-zero modes of the compact
Stueckelberg fields localize the dynamical gauge field to a flat connection.
The deformed torus partition function therefore takes the form
\begin{equation}
\begin{aligned}
    Z_{\lambda,C}[A_{\rm flat}]
    =
    \frac{|\det\mathsf M_{\lambda,C}|}{(2\pi)^2}
    \prod_{a=1}^{2}
    \int_{-\infty}^{\infty}
    d\widetilde\theta^a\,d\widetilde\nu^a\,
    \mathcal K_{\lambda,C}^{(1)}
    (\theta,\nu;\widetilde\theta,\widetilde\nu)\,
    Z_0[B_{\rm flat}] .
\end{aligned}
    \label{eq:intro-torus-flat-transform}
\end{equation}
Here \(A_{\rm flat}\) has lifted holonomies \((\theta^a,\nu^a)\), while
\(B_{\rm flat}\) has lifted holonomies
\((\widetilde\theta^a,\widetilde\nu^a)\).  All dependence on the seed theory is
contained in \(Z_0[B_{\rm flat}]\); the remaining factor is a universal kernel
determined by the deformation parameter and the anomaly representative.  The
same localization argument applies on a genus-\(g\) surface, where one obtains
a kernel \(\mathcal K_{\lambda,C}^{(g)}\) acting on the \(2g\) harmonic periods,
with only background-independent normalization factors left scheme dependent.

We illustrate the construction first with the compact boson, equivalently the
Abelian \(U(1)\) WZW model \cite{Witten:1991mm}.  In this case the
flat-background holonomy integral can be evaluated explicitly.  At vanishing
background gauge fields, the deformation only changes the radius parameter from
\(k\) to an effective parameter \(K_\lambda\).  Once background holonomies are
included, however, the full partition function is not obtained by the replacement
\(k\to K_\lambda\) throughout the undeformed answer: the spectrum is deformed in
the expected double-current way, while contact terms and spectral-flow data
remain controlled by the original anomaly coefficient.  We then consider
anomaly-free massive examples, namely massive complex bosons and massive Dirac
fermions.  Although their local deformed actions are different, the
flat-background kernel gives the same spectral statement: the deformed
finite-volume spectrum is the undeformed twisted spectrum evaluated at
charge-dependent twists. For the Dirac example this gives the Federbush model, whose
rapidity-independent mutual scattering phase leads to the same charge-dependent twist shifts.

Finally, we describe the extension beyond Abelian double-current
deformations.  In the non-Abelian case the deformation is not specified by the
symmetry algebra alone: one must also choose a Lie-algebra two-cocycle
\(\omega\), or equivalently, when it is non-degenerate on its image, the
corresponding homogeneous Yang-Baxter \(R\)-matrix.  This is the algebraic
data that appears in Deformed T-dual Models, which are equivalent to
homogeneous Yang-Baxter deformations in the appropriate non-degenerate case
\cite{Borsato:2016pas,Borsato:2017qsx}.  In our path integral formulation the
Abelian topological kernel is replaced by its Yang-Baxter analogue
\(\int_\Sigma \omega(\Theta\wedge\Theta)\), with the necessary global
\(G_\omega\)-reduction understood.  In the anomalous case this kernel is
supplemented by the same parallel transport in the anomaly line bundle as in
the Abelian construction.  Since the endpoint law of the transport is fixed
entirely by the finite anomaly, the resulting deformed partition function is
again a section of the same anomaly line bundle as the seed theory.

The paper is organized as follows.  In Section
\ref{sec:abelian-current-current-transforms} we give the path integral
formulation of Abelian double-current deformations, first in the anomaly-free
case and then in the presence of the anomaly, and explain its
current-current interpretation.  In
Section \ref{sec:partition-functions} we evaluate the flat-background path
integral on the torus and on a higher-genus Riemann surface.  In Section
\ref{sec:examples} we apply the general formulae to the compact boson and to two
massive anomaly-free QFTs, namely massive complex scalars and massive Dirac
fermions.  In Section \ref{sec:yang-baxter-generalization} we formulate the
non-Abelian and Yang-Baxter generalization. 

\setcounter{section}{1}
\section{Path integral formulation of Abelian double-current deformations}
\label{sec:abelian-current-current-transforms}

In this section we set up the path integral that defines the double-current deformation used throughout the paper; for general background on
current-current deformations see \cite{Borsato:2023dis}.  In the anomaly-free
case this is the construction of \cite{Dubovsky:2023lza}, written with a
dynamical \(U(1)^2\) gauge field and compact Stueckelberg fields so that small
and large auxiliary gauge transformations are treated uniformly.  For
anomalous seed theories, the same  topological kernel is retained, but
the auxiliary gauging must be made compatible with the anomaly line bundle of the background field theory.  This is achieved by inserting a Stueckelberg
anomaly factor and a parallel-transport factor in the anomaly line bundle.  The
anomaly is thus fixed input data, and the deformed generating functional has
the same finite background anomaly as the seed theory.

\subsection{Anomaly-free path integral formulation}
\label{subsec:anomaly-free-transform}

Let the seed theory be a two-dimensional Euclidean QFT on \(\Sigma\), with
fields \(\Phi\) and action \(S_0[\Phi]\).  The theory has two commuting
 \(U(1)\) global symmetries, labelled by \(a=1,2\), and \(\epsilon_{12}=1\).  In this subsection the corresponding
background coupling is non-anomalous, so the theory can be minimally coupled to
background gauge fields \(B^a\).  For a \(U(1)^2\)-valued gauge transformation
\(U^a:\Sigma\to U(1)\) we define
\begin{equation}
    \vartheta^a(U)
    \equiv
    -i\,(U^a)^{-1}dU^a .
    \label{eq:theta-U-definition}
\end{equation}
The non-anomalous coupling means that the action and the functional measure are
invariant under
\begin{equation}
    B^a\mapsto B^a-\vartheta^a(U),
    \qquad
    \Phi\mapsto U^{q[\Phi]}\Phi .
    \label{eq:anomaly-free-background-symmetry}
\end{equation}
Here \(U^{q[\Phi]}\equiv\prod_a(U^a)^{q_a[\Phi]}\), where \(q_a[\Phi]\) is the
charge of \(\Phi\) under the \(a\)-th \(U(1)\).
The conserved currents are defined by the response to the background fields,
\begin{equation}
    \delta_B S_0[\Phi|B]
    =
    -\int_\Sigma \delta B^a\wedge *J^a(B).
    \label{eq:undeformed-current-definition}
\end{equation}
At \(B=0\) these reduce to the conserved Noether currents of the original
theory.

The deformed theory is defined by promoting the background gauge fields in the
seed functional to dynamical gauge fields and adding the topological term.
The external background field is a pair of \(U(1)\) gauge fields \(A^a\) on principal
\(U(1)\) bundles \(P_A^a\to\Sigma\).  In the continuous
\(\lambda\)-deformation used in this paper, the dynamical gauge field \(B^a\) is
integrated only over bundles \(P_B^a\) in the same topological class,
\begin{equation}
    c_1(P_B^a)=c_1(P_A^a).
    \label{eq:same-class-sector}
\end{equation}
The compact Stueckelberg field \(h^a\) is a trivialization of the relative
bundle \(P_A^a\otimes(P_B^a)^{-1}\); after choosing such a trivialization it is
represented by a map \(\Sigma\to U(1)\).  Define
\begin{equation}
    B_h^a
    =
    B^a-\vartheta^a(h),
    \qquad
    \Theta^a
    =
    A^a-B_h^a
    =
    A^a-B^a+\vartheta^a(h).
    \label{eq:Bh-Theta}
\end{equation}
Because of \eqref{eq:same-class-sector}, \(\Theta^a\) is an ordinary
globally defined one-form on \(\Sigma\).  If \(A^a\) and \(B^a\) are allowed to
belong to different bundle classes, the same formula must instead be interpreted
as a differential-cohomology pairing \cite{Davighi:2020vcm}.  The all-sector
version would quantize the coefficient of this topological term, or
equivalently quantize \(\lambda\).  In this paper \(\lambda\) is kept continuous, and the
deformation is restricted to the same-class sector.

The anomaly-free path integral transform is
\begin{equation}
\begin{aligned}
    Z_\lambda[A]
    =
    \int
    \frac{[D\Phi\,DB\,Dh]}{\mathrm{Vol}(\mathcal G)}
    \exp\!\left[
    -S_0[\Phi|B]
    -
    \frac{i}{4\lambda}
    \int_\Sigma \epsilon_{ab}\Theta^a\wedge\Theta^b
    \right],
\end{aligned}
    \label{eq:anomaly-free-transform}
\end{equation}
where \(\mathcal G=\mathrm{Map}(\Sigma,U(1)^2)\).  The denominator implements the
auxiliary gauge redundancy
\begin{equation}
    B^a\mapsto B^a-\vartheta^a(U),
    \qquad
    h^a\mapsto (U^a)^{-1}h^a,
    \qquad
    \Phi\mapsto U^{q[\Phi]}\Phi .
    \label{eq:anomaly-free-dynamical-gauge-symmetry}
\end{equation}
The combination \(\Theta^a\) is invariant, and the matter factor is invariant by
assumption.  The deformed functional also has the expected background gauge
covariance:
under
\begin{equation}
    A^a\mapsto A^a-\vartheta^a(V),
    \qquad
    h^a\mapsto V^a h^a,
    \label{eq:anomaly-free-background-gauge-symmetry}
\end{equation}
the one-forms \(\Theta^a\) are unchanged.

The relation with a double-current deformation is seen at the saddle point of
the auxiliary sector.  In unitary gauge \(h^a=1\), variation with
respect to the dynamical gauge field gives
\begin{equation}
    *J^a(B_*)
    =
    \frac{i}{2\lambda}\,
    \epsilon_{ab}(B_*^b-A^b).
    \label{eq:anomaly-free-B-saddle}
\end{equation}
Here \(B_*^a\) is the dynamical gauge-field saddle and \(J^a(B)\) is the current
defined by \eqref{eq:undeformed-current-definition}.  The current of the
deformed theory is defined by the on-shell response to the background gauge
field,
\begin{equation}
    \delta_A S_\lambda[\Phi|A]
    =
    -\int_\Sigma \delta A^a\wedge *J_\lambda^a(A),
    \qquad
    J_\lambda^a(A)=J^a(B_*).
    \label{eq:anomaly-free-deformed-current}
\end{equation}
The saddle equation removes the implicit variation of \(B_*\).  Differentiating
the explicit quadratic topological term with respect to \(\lambda\) therefore gives
\begin{equation}
    \frac{d}{d\lambda}S_\lambda[\Phi|A]
    =
    i\epsilon_{ab}
    \int_\Sigma
    J_\lambda^a(A)\wedge J_\lambda^b(A).
    \label{eq:anomaly-free-current-current-flow}
\end{equation}
Thus the path integral transform is a quantum definition of the finite
deformation whose classical flow is generated by the antisymmetric bilinear of
the two currents.  The anomalous formulation below keeps the same topological
gauging term and adds the anomaly-line transport and Stueckelberg compensator
required by the finite anomaly.

\subsection{Anomalous path integral formulation}
\label{subsec:anomalous-transform}

Now suppose that the \(U(1)^2\) background gauge-field coupling of the seed
theory has a finite anomaly.  The anomaly is part of the input data and we assume the anomaly is kept independent
of \(\lambda\).  This is important globally: a \(\lambda\)-dependent anomaly can change the quantization conditions of the background theory,
rather than only deforming its dynamics.  The undeformed generating functional
is
\begin{equation}
    Z_0[B]
    =
    \int[D\Phi]\exp\!\left[-S_0[\Phi|B]\right].
    \label{eq:anomalous-undeformed-generating-functional}
\end{equation}
Under the finite background gauge transformation
\(B^a\mapsto B^a-\vartheta^a(U)\), we assume
\begin{equation}
    Z_0[B-\vartheta(U)]
    =
    \exp\!\left[-\mathcal I_{\rm anom}[U,B]\right]Z_0[B].
    \label{eq:finite-anomaly-functional}
\end{equation}
Here \(\mathcal I_{\rm anom}[U,B]\) is the finite anomaly functional; \(U\) and
\(B\) denote the full \(U(1)^2\) collection.  It satisfies the
Wess-Zumino consistency condition
\cite{Wess:1971yu}
\begin{equation}
    \mathcal I_{\rm anom}[UV,B]
    =
    \mathcal I_{\rm anom}[U,B]
    +
    \mathcal I_{\rm anom}[V,B-\vartheta(U)] .
    \label{eq:finite-anomaly-wz-consistency}
\end{equation}
This finite cocycle contains both perturbative and global anomaly data. The anomalous partition function is not an ordinary function on
the space of background gauge fields, but a section of an anomaly line bundle
\cite{Bismut:1986wr,Freed:1992vw,Freed:2014iua}.

The auxiliary gauge quotient in the anomaly-free definition is no longer
well-defined by itself, because gauge-related dynamical gauge-field
configurations have weights that differ by \(\mathcal I_{\rm anom}\).  We keep the same-class
sector \eqref{eq:same-class-sector} and the definitions of \(B_h^a\) and
\(\Theta^a\) in \eqref{eq:Bh-Theta}.  Thus \(\Theta^a\) is again an ordinary
globally defined one-form.  If \(A^a\) and \(B^a\) are not in the same bundle
class, the topological term must be lifted to a differential-cohomology
pairing; for continuous \(\lambda\) we do not include those sectors.  In the anomalous case
one must also insert a transport factor in the anomaly line bundle.

Let \(T[A,B]\) be a choice of transport between two connections in this component
of background-field space, viewed as a local representative of parallel transport
in the anomaly line \cite{Bismut:1986wr}.  Its endpoint gauge law is
\begin{equation}
    T[A-\vartheta(V),B-\vartheta(V)]
    =
    \exp\!\left[
    -\mathcal I_{\rm anom}[V,A]
    +
    \mathcal I_{\rm anom}[V,B]
    \right]T[A,B],
    \label{eq:transport-endpoint-law}
\end{equation}
for every finite gauge transformation \(V\), together with the diagonal
normalization
\begin{equation}
    T[A,A]=1 .
    \label{eq:transport-diagonal-normalization}
\end{equation}
A convenient local construction uses a connection one-form \(\Omega\) on the
anomaly line over the same component of background-field space.  Let
\(R_V(B)=B-\vartheta(V)\), and let \(\delta_B\) denote the exterior derivative on
this space.  The required equivariance is
\begin{equation}
    R_V^*\Omega-\Omega
    =
    \delta_B\mathcal I_{\rm anom}[V,B].
    \label{eq:Omega-equivariance}
\end{equation}
Given a path \(\gamma\) from \(B\) to \(A\), a compatible representative is then
\begin{equation}
    T_\gamma[A,B]
    =
    \exp\!\left[-\int_\gamma\Omega\right],
    \label{eq:transport-from-Omega}
\end{equation}
provided the path prescription is gauge covariant,
\(\gamma[A-\vartheta(V),B-\vartheta(V)]=R_V\gamma[A,B]\).  Equation
\eqref{eq:Omega-equivariance} then gives
\eqref{eq:transport-endpoint-law}.  In general this representative can depend
on the path.  It is locally path independent when \(\Omega\) is closed, and
globally path independent only when the corresponding anomaly-line holonomies
are trivial.

The finite anomaly representative is scheme dependent.  Adding a local
counterterm \(L[B]\), \(Z_0[B]\mapsto e^{-L[B]}Z_0[B]\), shifts it as
\begin{equation}
    \mathcal I_{\rm anom}[U,B]
    \mapsto
    \mathcal I_{\rm anom}[U,B]
    +
    L[B-\vartheta(U)]-L[B].
    \label{eq:anomaly-scheme-shift}
\end{equation}
The transport changes accordingly,
\begin{equation}
    T[A,B]
    \mapsto
    e^{-L[A]+L[B]}T[A,B],
    \label{eq:transport-scheme-shift}
\end{equation}
so that the endpoint law and the diagonal normalization are preserved in the
new scheme.  Even after the representative has been fixed, the transport is not
unique.  If \(T'\) obeys the same endpoint law and normalization, then
\begin{equation}
    \Lambda_T[A,B]\equiv \frac{T'[A,B]}{T[A,B]}
    \label{eq:transport-ratio}
\end{equation}
satisfies
\begin{equation}
    \Lambda_T[A-\vartheta(V),B-\vartheta(V)]
    =
    \Lambda_T[A,B],
    \qquad
    \Lambda_T[A,A]=1 .
    \label{eq:transport-nonuniqueness}
\end{equation}
Such a factor leaves the anomaly and the undeformed limit unchanged, but it is
part of the finite-\(\lambda\) definition of the path integral transform.

The anomalous double-current deformation is defined by the path integral
transform
\begin{equation}
\begin{aligned}
    Z_\lambda[A]
    &=
    \int\frac{[D\Phi\,DB\,Dh]}{\mathrm{Vol}(\mathcal G)}\,
    T[A,B_h]
    \exp\!\left[-S_0[\Phi|B]-\mathcal I_{\rm anom}[h,B]
    -\frac{i}{4\lambda}\int_\Sigma\epsilon_{ab}\Theta^a\wedge\Theta^b\right].
\end{aligned}
    \label{eq:anomalous-transform}
\end{equation}
The quotient is by the same auxiliary gauge action as in
\eqref{eq:anomaly-free-dynamical-gauge-symmetry}.  The combination \(B_h^a\),
and hence \(\Theta^a\), is invariant.  The anomalous variation of the matter
functional is cancelled by the Stueckelberg compensator.
Indeed, \eqref{eq:finite-anomaly-wz-consistency} gives
\begin{equation}
    \mathcal I_{\rm anom}[U^{-1}h,B-\vartheta(U)]
    =
    \mathcal I_{\rm anom}[h,B]
    -
    \mathcal I_{\rm anom}[U,B].
    \label{eq:anomalous-dynamical-cancellation}
\end{equation}
This shows that the integrand of \eqref{eq:anomalous-transform} descends to the
auxiliary gauge quotient. In fact, this action is equivalent to 
\begin{equation}
    Z_{\lambda}[A]=\int [ DB]T[A,B]Z_0[B]\exp\left[-\frac{i}{4\lambda}\epsilon_{ab}(A^a-B^a)\wedge(A^b-B^b)\right]
\end{equation}
To keep the Stueckelberg field manifest, we keep the formulation \eqref{eq:anomalous-transform}.

The same definition fixes the background gauge covariance of the deformed
functional.  Under \eqref{eq:anomaly-free-background-gauge-symmetry} one has
\(B_{Vh}=B_h-\vartheta(V)\), while \(\Theta^a\) is unchanged.  Using
\eqref{eq:transport-endpoint-law} and
\begin{equation}
    \mathcal I_{\rm anom}[Vh,B]
    =
    \mathcal I_{\rm anom}[h,B]
    +
    \mathcal I_{\rm anom}[V,B_h],
    \label{eq:anomalous-background-cocycle}
\end{equation}
inside the path integral gives
\begin{equation}
    Z_\lambda[A-\vartheta(V)]
    =
    \exp\!\left[-\mathcal I_{\rm anom}[V,A]\right]Z_\lambda[A].
    \label{eq:anomalous-transform-preserves-anomaly}
\end{equation}
Thus the deformed theory has the same finite background anomaly as the seed
theory.  If \(\mathcal I_{\rm anom}=0\), the minimal transport is \(T=1\) and
\eqref{eq:anomalous-transform} reduces to the anomaly-free deformation
\eqref{eq:anomaly-free-transform}.

\paragraph{Example.}
A useful local representative of an Abelian anomaly is specified by a constant
matrix \(C_{ab}\), as in \cite{Dubovsky:2023lza}.  For a finite \(U(1)^2\)
gauge transformation \(Q^a\), define
\begin{equation}
    \mathcal W_C[Q,B]
    =
    C_{ab}
    \int_Y
    \vartheta^a(\widehat Q)\wedge d\widehat{\mathcal B}^b ,
    \qquad
    \partial Y=\Sigma ,
    \qquad
    \iota_\Sigma^*\widehat{\mathcal B}^a=B^a,
    \qquad
    \iota_\Sigma^*\widehat Q^a=Q^a .
    \label{eq:C-anomaly-WZ-functional}
\end{equation}
Here hatted fields denote three-dimensional extensions to \(Y\).  The
exponential of \(\mathcal W_C\) is independent of the extensions subject to the
usual integrality condition on the finite anomaly levels.  With the convention
\begin{equation}
    \mathcal I_{\rm anom}[U,B]
    =
    -
    \mathcal W_C[U,B],
    \label{eq:C-anomaly-functional}
\end{equation}
the undeformed generating functional obeys
\eqref{eq:finite-anomaly-functional}.  The local connection one-form
\begin{equation}
    \Omega_C(B)
    =
    -
    C_{ab}\int_\Sigma B^a\wedge\delta B^b
    \label{eq:C-anomaly-Omega}
\end{equation}
is compatible with \eqref{eq:Omega-equivariance}.  Indeed,
\begin{equation}
    \delta_B\mathcal W_C[V,B]
    =
    -
    C_{ab}\int_\Sigma\vartheta^a(V)\wedge\delta B^b,
    \qquad
    R_V^*\Omega_C-\Omega_C
    =
    C_{ab}\int_\Sigma\vartheta^a(V)\wedge\delta B^b,
    \label{eq:C-anomaly-Omega-check}
\end{equation}
which equals \(\delta_B\mathcal I_{\rm anom}[V,B]\) because
\(\mathcal I_{\rm anom}=-\mathcal W_C\).

Equivalently, let \(\mathfrak A_C\) denote the three-dimensional invertible
anomaly theory.  On a closed three-manifold \(M\), equipped with a
three-dimensional background connection \(\widehat{\mathcal A}^a\), a local
representative of its background response is
\begin{equation}
    Z_C(M,\widehat{\mathcal A})
    =
    \exp\!\left[
    -
    C_{ab}\int_M
    \widehat{\mathcal A}^a\wedge d\widehat{\mathcal A}^b
    \right].
    \label{eq:C-anomaly-theory}
\end{equation}
In particular, for a three-dimensional connection \(\widehat{\mathcal A}^a\) on
\(\Sigma\times I\) whose boundary pullbacks obey
\begin{equation}
    \iota_0^*\widehat{\mathcal A}^a=B^a,
    \qquad
    \iota_1^*\widehat{\mathcal A}^a=A^a,
    \label{eq:C-cylinder-boundary}
\end{equation}
the Abelian anomaly theory gives the anomaly-line transport
\begin{equation}
    T_C[A,B;\widehat{\mathcal A}]
    :=
    Z_C(\Sigma\times I,\widehat{\mathcal A}) .
    \label{eq:C-cylinder-amplitude}
\end{equation}
Choosing temporal gauge for \(\widehat{\mathcal A}^a\) represents the
interpolation by a path \(\gamma_s^a\) of two-dimensional connections, with
\(\gamma_0^a=B^a\) and \(\gamma_1^a=A^a\).  For such a representative, this
gives
\begin{equation}
    T_{C,\gamma}[A,B]
    =
    \exp\!\left[
    C_{ab}\int_0^1 ds
    \int_\Sigma \gamma_s^a\wedge\dot\gamma_s^b
    \right].
    \label{eq:C-anomaly-transport}
\end{equation}
For any gauge-covariant path prescription this obeys
\eqref{eq:transport-endpoint-law}.  For the straight path
\begin{equation}
    \gamma_s^a=B^a+s(A^a-B^a),
    \label{eq:C-anomaly-straight-path}
\end{equation}
this representative becomes
\begin{equation}
    T_C[A,B]
    =
    \exp\!\left[
    C_{ab}\int_\Sigma B^a\wedge(A^b-B^b)
    +
    \frac12 C_{ab}\int_\Sigma
    (A^a-B^a)\wedge(A^b-B^b)
    \right].
    \label{eq:C-anomaly-transport-straight}
\end{equation}
Substituting \eqref{eq:C-anomaly-functional} and
\eqref{eq:C-anomaly-transport-straight} into
\eqref{eq:anomalous-transform}, with \(B\) replaced by \(B_h\) in the
transport, gives
\begin{equation}
\begin{aligned}
    Z_{\lambda,C}[A]
    &=
    \int
    \frac{[D\Phi\,DB\,Dh]}{\mathrm{Vol}(\mathcal G)}
    \exp\!\Bigg[
    -S_0[\Phi|B]
    +
    \mathcal W_C[h,B]
    \\
    &\hspace{2.2cm}
    -
    \frac{i}{4\lambda}
    \int_\Sigma\epsilon_{ab}\Theta^a\wedge\Theta^b
    +
    C_{ab}\int_\Sigma A^a\wedge\Theta^b
    -
    \frac12 C_{ab}\int_\Sigma\Theta^a\wedge\Theta^b
    \Bigg].
\end{aligned}
    \label{eq:C-anomaly-explicit-transform}
\end{equation}
When the background and dynamical gauge fields are not represented by ordinary
global one-forms, the terms in
\eqref{eq:C-anomaly-transport-straight} and
\eqref{eq:C-anomaly-explicit-transform} are understood as local representatives
of the corresponding anomaly-line pairing.

\subsection{Current-current interpretation}
\label{subsec:current-current-interpretation}

The anomalous path integral transform has the same explicit \(\lambda\)-flow as
the anomaly-free transform.  The anomaly-dependent factors \(T[A,B_h]\) and
\(\exp[-\mathcal I_{\rm anom}[h,B]]\) are independent of \(\lambda\), while
the \(\lambda\)-dependence is carried by the quadratic topological term.
Therefore
\begin{equation}
    \partial_\lambda \log Z_\lambda[A]
    =
    \frac{i}{4\lambda^2}
    \left\langle
    \int_\Sigma\epsilon_{ab}\Theta^a\wedge\Theta^b
    \right\rangle_{\lambda,A}.
    \label{eq:anomalous-lambda-flow}
\end{equation}
This is an exact identity for the background-field dependent partition function.  The
classical current-current interpretation is its saddle-point form.  By
Wess-Zumino consistency, the compensator vanishes at \(h^a=1\).  Hence, in
unitary gauge, the classical action entering the deformed saddle is
\begin{equation}
    S_T[A,B]=-\log T[A,B],
    \qquad
    S_{\rm cl}
    =
    S_0[\Phi|B]
    +
    S_T[A,B]
    +
    \frac{i}{4\lambda}
    \int_\Sigma\epsilon_{ab}(A^a-B^a)\wedge(A^b-B^b).
    \label{eq:anomalous-saddle-action}
\end{equation}
The endpoint responses of the anomaly-line transport and the seed-theory
current are defined by
\begin{equation}
    \delta S_T
    =
    \int_\Sigma\delta A^a\wedge\Pi^A_a[A,B]
    +
    \int_\Sigma\delta B^a\wedge\Pi^B_a[A,B],
    \qquad
    \delta_BS_0[\Phi|B]
    =
    -\int_\Sigma\delta B^a\wedge *J^a(B).
    \label{eq:transport-responses}
\end{equation}
Let \(B_*^a\) denote the dynamical gauge-field saddle.  The one-form current
associated with the quadratic topological term is
\begin{equation}
    *H_\lambda^a(A)
    \equiv
    -
    \frac{i}{2\lambda}\epsilon_{ab}(A^b-B_*^b).
    \label{eq:kernel-current}
\end{equation}
whereas the physical current of the deformed saddle is defined by its response to
the background gauge field,
\(\delta_A S_\lambda[\Phi|A]=-\int_\Sigma\delta A^a\wedge
*J_\lambda^a(A)\).  The \(B\)-equation and the on-shell response to the
background field give
\begin{equation}
\begin{aligned}
    *H_\lambda^a(A)
    &=
    *J^a(B_*)-\Pi^B_a[A,B_*],
    \\
    *J_\lambda^a(A)
    &=
    *J^a(B_*)-\Pi^B_a[A,B_*]-\Pi^A_a[A,B_*].
\end{aligned}
    \label{eq:anomalous-current-relations}
\end{equation}
Thus the classical flow is governed by the same quadratic term as in the
non-anomalous theory, but the current appearing in that term is
not generically the physical background-field current.  Their difference is the
endpoint response of the anomaly-line transport.

The same endpoint response also fixes the Ward identity.  The infinitesimal
form of the endpoint law \eqref{eq:transport-endpoint-law} is
\begin{equation}
    d\!\left(\Pi^A_a[A,B]+\Pi^B_a[A,B]\right)
    =
    \mathfrak a_a[A]-\mathfrak a_a[B],
    \qquad
    \mathcal I_{\rm anom}[e^{i\varepsilon},A]
    =
    \int_\Sigma\varepsilon^a\,\mathfrak a_a[A]+O(\varepsilon^2).
    \label{eq:endpoint-response-ward}
\end{equation}
At the matter saddle \(d*J^a(B)=-\mathfrak a_a[B]\).  Combining this relation
with \eqref{eq:anomalous-current-relations} gives
\begin{equation}
    d*J_\lambda^a(A)=-\mathfrak a_a[A].
    \label{eq:deformed-current-anomalous-ward}
\end{equation}
The transport formulation therefore preserves the finite anomaly without
turning the anomalous background-field current into an ordinary conserved
current.  At zero background the deformed current is conserved only when the chosen anomaly
representative has \(\mathfrak a_a[0]=0\).

In a local patch of background-field space the transport can be represented by a
one-form connection
\begin{equation}
    \Omega(B)
    =
    \int_\Sigma\delta B^a\wedge\omega_a[B],
    \label{eq:local-Omega-representative}
\end{equation}
with \(\delta B^a\) the field-space one-form.  If this representative is
locally closed, the transport is locally path independent and the endpoint
responses reduce to
\begin{equation}
    \Pi^A_a[A,B]=\omega_a[A],
    \qquad
    \Pi^B_a[A,B]=-\omega_a[B].
    \label{eq:closed-Omega-responses}
\end{equation}
In this representative,
\begin{equation}
    *J_\lambda^a(A)
    =
    *J^a(B_*)+\omega_a[B_*]-\omega_a[A],
    \qquad
    *H_\lambda^a(A)
    =
    *J^a(B_*)+\omega_a[B_*].
    \label{eq:closed-Omega-current-redefinition}
\end{equation}
The saddle-point flow is therefore
\begin{equation}
    \frac{d}{d\lambda}S_\lambda[\Phi|A]
    =
    i\epsilon_{ab}
    \int_\Sigma H_\lambda^a(A)\wedge H_\lambda^b(A).
    \label{eq:anomalous-classical-current-current-flow}
\end{equation}
No additional \(\lambda\)-flow is generated by the anomaly compensator; its
effect is the current redefinition in \eqref{eq:anomalous-current-relations}.
If, in addition, the representative is chosen so that \(\omega_a[0]=0\), then
at zero background \(H_\lambda^a\) coincides with the physical deformed current
and the background-free deformation reduces to the ordinary double-current flow.

\section{Flat-background partition functions}
\label{sec:partition-functions}

In this section we evaluate the path integral transform defined above for flat
background gauge fields on compact Euclidean Riemann surfaces.  Torus partition
functions are particularly useful, since
their dependence on twists and chemical potentials encodes finite-volume
spectra and charge assignments
\cite{Datta:2018thy,Aharony:2018bad,Aharony:2018ics}.  From the
path integral viewpoint, the same data are captured by universal holonomy
kernels of the type appearing in topological-gauging constructions
\cite{Dubovsky:2018bmo,Aguilera-Damia:2019tpe,Dubovsky:2023lza,Borsato:2023dis}.

Throughout this section the dynamical gauge field is integrated in the same
topological sector as the background gauge field.  For flat backgrounds, the
non-zero modes of the compact Stueckelberg fields impose the flatness of the
dynamical gauge field.  After the local gauge-fixing determinants are absorbed
into the normalization, the path integral reduces to a finite-dimensional
integral over real lifts of the holonomies of the localized flat connection. On the torus, the \(A\)-cycle holonomies specify twisted sectors, while the
\(B\)-cycle holonomies play the role of chemical potentials.  We first derive
the genus-one kernel and explain its Hamiltonian interpretation.  We then give
the higher-genus generalization in a symplectic basis of harmonic one-forms.

\subsection{Torus flat-background partition function}
\label{subsec:torus-flat-background-transform}

We begin with \(T^2\), where the flat holonomies have a direct interpretation as
twists and chemical potentials.  The goal is to evaluate the path integral
transform \eqref{eq:anomalous-transform} as a finite-dimensional definition of
the deformed flat-background partition function in terms of the undeformed one.

Let
\begin{equation}
    z\sim z+2\pi\sim z+2\pi\tau,
    \qquad
    \tau=\tau_1+i\tau_2,
    \qquad
    \tau_2>0 ,
    \label{eq:torus-identification}
\end{equation}
with \(z=\sigma^1+i\sigma^2\).  We use the convention
\begin{equation}
    d^2z=i\,dz\wedge d\bar z .
    \label{eq:complex-measure-convention}
\end{equation}
The two cycles are generated by
\begin{equation}
    (\sigma^1,\sigma^2)\mapsto(\sigma^1+2\pi,\sigma^2),
    \qquad
    (\sigma^1,\sigma^2)\mapsto
    (\sigma^1+2\pi\tau_1,\sigma^2+2\pi\tau_2).
    \label{eq:torus-cycles}
\end{equation}
The harmonic one-forms dual to these cycles are chosen as
\begin{equation}
    \rho=
    \frac{d\sigma^1}{2\pi}
    -
    \frac{\tau_1}{\tau_2}\frac{d\sigma^2}{2\pi},
    \qquad
    \sigma=
    \frac{d\sigma^2}{2\pi\tau_2},
    \label{eq:torus-dual-one-forms}
\end{equation}
so that \(\oint_A\rho=\oint_B\sigma=1\),
\(\oint_A\sigma=\oint_B\rho=0\), and
\(\int_{T^2}\rho\wedge\sigma=1\).  A flat background gauge field is written as
\begin{equation}
    A_{\rm flat}^a
    =
    \theta^a\rho+\nu^a\sigma
    =
    \frac{\theta^a}{2\pi}\,d\sigma^1
    +
    \frac{\nu^a-\tau_1\theta^a}{2\pi\tau_2}\,d\sigma^2,
    \qquad
    a=1,2 .
    \label{eq:torus-flat-background}
\end{equation}
Thus \(\theta^a=\oint_A A_{\rm flat}^a\) and
\(\nu^a=\oint_B A_{\rm flat}^a\).  Large gauge
transformations shift
\begin{equation}
    \theta^a\mapsto\theta^a-2\pi m^a,
    \qquad
    \nu^a\mapsto\nu^a-2\pi n^a,
    \qquad
    m^a,n^a\in\mathbb Z .
    \label{eq:torus-large-gauge-shifts}
\end{equation}
Because the background gauge field is flat, the same-class prescription of
\S\ref{subsec:anomalous-transform} allows us to use globally defined
representatives for \(A_{\rm flat}^a\), \(B^a\), and the relative one-form
\(\Theta^a\).

For the \(C_{ab}\) anomaly representative of
Section \ref{subsec:anomalous-transform}, the deformed flat-background
partition function is
\begin{equation}
\begin{aligned}
    Z_{\lambda,C}[A_{\rm flat}]
    &=
    \int\frac{[DB^1\,DB^2\,Dh^1\,Dh^2]}{\mathrm{Vol}(\mathcal G)}\,
    Z_0[B]\exp\!\Bigg[\mathcal W_C[h,B]
    -\frac{i}{4\lambda}\int_{T^2}\epsilon_{ab}\Theta^a\wedge\Theta^b
    \\
    &\hspace{34mm}
    +C_{ab}\int_{T^2}A_{\rm flat}^a\wedge\Theta^b
    -\frac12 C_{ab}\int_{T^2}\Theta^a\wedge\Theta^b\Bigg],
\end{aligned}
    \label{eq:torus-dynamical-path integral}
\end{equation}
where
\begin{equation}
    \Theta^a=A_{\rm flat}^a-B^a+\vartheta(h^a).
    \label{eq:torus-Theta}
\end{equation}
All seed-theory dependence is contained in \(Z_0[B]\).  The universal
dynamical gauge-field and Stueckelberg factors are evaluated in Appendix
\ref{subsec:app-torus-path integral}.  We use a gauge slice in which small gauge
transformations remove the exact part of \(B^a\), while pure large
transformations remove the winding of \(h^a\):
\begin{equation}
    B^a=B^a_{\rm harm}+*d\psi^a,
    \qquad
    \vartheta(h^a)=dX^a ,
    \label{eq:torus-gauge-slice}
\end{equation}
where \(X^a\) is a single-valued lift of the remaining compact field and the
constant modes of \(\psi^a\) are omitted.

On this slice the dependence on the single-valued fields \(X^a\) is linear:
\begin{equation}
    \mathcal E_X
    =
    \int_{T^2} dX^c\wedge
    \left[
    \frac{i}{2\lambda}\epsilon_{ac}(A_{\rm flat}^a-B^a)
    -
    \frac12(C_{ac}+C_{ca})(A_{\rm flat}^a+B^a)
    \right].
    \label{eq:torus-X-linear-coupling}
\end{equation}
For flat \(A_{\rm flat}\), the non-zero modes of the Stueckelberg fields impose
\begin{equation}
    \mathsf M_{\lambda,C}
    \begin{pmatrix}
        dB^1 \\ dB^2
    \end{pmatrix}
    =
    0,
    \qquad
    (\mathsf M_{\lambda,C})_{ca}
    =
    -
    \frac{1}{2\lambda}\epsilon_{ac}
    +
    \frac{i}{2}(C_{ac}+C_{ca}) .
    \label{eq:torus-flatness-matrix}
\end{equation}
In components,
\begin{equation}
    \mathsf M_{\lambda,C}
    =
    \begin{pmatrix}
        iC_{11}
        &
        \frac{1}{2\lambda}+\frac{i}{2}(C_{12}+C_{21})
        \\
        -\frac{1}{2\lambda}+\frac{i}{2}(C_{12}+C_{21})
        &
        iC_{22}
    \end{pmatrix},
    \label{eq:torus-flatness-matrix-explicit}
\end{equation}
with
\begin{equation}
    \Delta_{\lambda,C}
    \equiv
    \det\mathsf M_{\lambda,C}
    =
    \frac{1}{4\lambda^2}
    +
    \frac14(C_{12}+C_{21})^2
    -
    C_{11}C_{22}.
    \label{eq:torus-flatness-determinant}
\end{equation}
Away from the singular locus \(\Delta_{\lambda,C}=0\), this constraint localizes
the dynamical gauge field to flat configurations:
\begin{equation}
    dB^1=dB^2=0 .
    \label{eq:torus-dynamical-flatness}
\end{equation}
The scalar Faddeev-Popov determinants from small gauge fixing cancel the scalar
determinants produced by the non-zero-mode Fourier transform over \(X^a\), as
shown in Appendix \ref{subsec:app-torus-path integral}.  The remaining
background-independent factor is fixed by the undeformed limit.  Thus the
remaining integration reduces to lifted real holonomies.  On the localized slice,
\(B^a=B_{\rm harm}^a\); we denote this harmonic flat connection by
\(B_{\rm flat}^a\):
\begin{equation}
    B_{\rm flat}^a
    =
    \widetilde\theta^a\rho+\widetilde\nu^a\sigma
    =
    \frac{\widetilde\theta^a}{2\pi}d\sigma^1
    +
    \frac{\widetilde\nu^a-\tau_1\widetilde\theta^a}
    {2\pi\tau_2}d\sigma^2,
    \qquad
    \widetilde\theta^a,\widetilde\nu^a\in\mathbb R .
    \label{eq:torus-flat-dynamical-field}
\end{equation}
For two flat one-forms
\(\alpha=\theta_\alpha\rho+\nu_\alpha\sigma\) and
\(\beta=\theta_\beta\rho+\nu_\beta\sigma\),
\begin{equation}
    \int_{T^2}\alpha\wedge\beta
    =
    \theta_\alpha\nu_\beta-\nu_\alpha\theta_\beta .
    \label{eq:torus-harmonic-wedge}
\end{equation}

Set
\begin{equation}
    \Delta\theta^a=\theta^a-\widetilde\theta^a,
    \qquad
    \Delta\nu^a=\nu^a-\widetilde\nu^a .
    \label{eq:torus-holonomy-differences}
\end{equation}
The resulting genus-one kernel is
\begin{equation}
\begin{aligned}
    \mathcal K_{\lambda,C}^{(1)}
    (\theta^a,\nu^a;\widetilde\theta^a,\widetilde\nu^a)
    =
    \exp\!\Big[
    C_{ab}(\widetilde\theta^a\Delta\nu^b-\widetilde\nu^a\Delta\theta^b)
    +\left(\frac12 C_{ab}-\frac{i}{4\lambda}\epsilon_{ab}\right)
    (\Delta\theta^a\Delta\nu^b-\Delta\nu^a\Delta\theta^b)\Big].
\end{aligned}
    \label{eq:torus-flat-kernel}
\end{equation}
The superscript \((1)\) indicates that this is the genus-one finite-dimensional
kernel.
With the normalization fixed below, the deformed partition function is
\begin{equation}
\begin{aligned}
    Z_{\lambda,C}[A_{\rm flat}]
    =
    \frac{|\det\mathsf M_{\lambda,C}|}{(2\pi)^2}
    \prod_{a=1}^{2}
    \int_{-\infty}^{\infty}
    d\widetilde\theta^a\,d\widetilde\nu^a\,
    Z_0[B_{\rm flat}]\,
    \mathcal K_{\lambda,C}^{(1)}
    (\theta^a,\nu^a;\widetilde\theta^a,\widetilde\nu^a).
\end{aligned}
    \label{eq:torus-flat-kernel-transform}
\end{equation}
The normalization is chosen so that the path integral transform becomes the
identity operation on flat-background partition functions as \(\lambda\to0\).  Indeed,
\(\det\mathsf M_{\lambda,C}=1/(4\lambda^2)+O(1)\), and the rapidly oscillating
topological phase becomes the Fourier representation of the delta function on
the four lifted holonomies,
\begin{equation}
    \frac{|\det\mathsf M_{\lambda,C}|}{(2\pi)^2}
    \mathcal K_{\lambda,C}^{(1)}
    \longrightarrow
    \prod_{a=1}^{2}
    \delta(\theta^a-\widetilde\theta^a)\,
    \delta(\nu^a-\widetilde\nu^a).
    \label{eq:torus-undeformed-limit}
\end{equation}
The \(C_{ab}\)-dependent part of the phase is trivial on the support of this
delta kernel, so \eqref{eq:torus-flat-kernel-transform} reduces to
\(Z_0[A_{\rm flat}]\).

The same background holonomies have the standard Hamiltonian interpretation.  The
\(A\)-cycle periods \(\theta^a\) specify the twisted Hilbert space, while the
\(B\)-cycle periods \(\nu^a\) are chemical potentials.  After separating local
background-field terms, the flat-background partition function can be written as
\begin{equation}
    Z_{\lambda,C}[A_{\rm flat}]
    =
    e^{-S_{{\rm ct},\lambda}(\theta^a,\nu^a)}
    {\rm Tr}_{\mathcal H_{\theta}}
    \exp\!\left[
    -2\pi\tau_2 H_\lambda(\theta)
    +2\pi i\tau_1 P_\lambda(\theta)
    +i\nu^a Q_{a,\lambda}(\theta)
    \right].
    \label{eq:torus-flat-background-trace}
\end{equation}
Here \(S_{{\rm ct},\lambda}\) is scheme dependent and contains the contact terms together with the local background-field terms generated by the deformation.
The spectrum and charge operators are read from the non-local part of the trace;
the contact term is fixed separately by the anomaly scheme.

\subsection{Higher-genus flat-background partition function}
\label{subsec:higher-genus-flat-background-transform}

We now extend the result to a compact Riemann surface \(\Sigma_g\).  Since the
non-zero-mode analysis is local, it is the same as on the torus: the dynamical
gauge field is localized to \(B_{\rm flat}^a\), and the remaining local
determinants are independent of the background field.  We therefore record only
the finite harmonic integral that defines the deformed flat-background partition function;
the gauge-fixing derivation is given in Appendix
\ref{subsec:app-higher-genus-derivation}.  As in Section
\ref{subsec:torus-flat-background-transform}, the dynamical and background gauge
fields are taken in the same bundle class, \(P_B^a\simeq P_A^a\).  For
flat \(U(1)\) connections on \(\Sigma_g\) this allows us, after choosing a
trivialization, to represent both \(A_{\rm flat}^a\) and \(B_{\rm flat}^a\) by
globally defined closed one-forms.

Choose a symplectic basis of one-cycles
\(\{A_I,B_I\}_{I=1}^g\).  Let \(\rho_I,\sigma_I\) be real harmonic one-forms
normalized by
\begin{equation}
    \int_{A_J}\rho_I=\delta_{IJ},
    \qquad
    \int_{B_J}\sigma_I=\delta_{IJ},
    \qquad
    \int_{A_J}\sigma_I=\int_{B_J}\rho_I=0,
    \label{eq:higher-genus-period-basis}
\end{equation}
and oriented so that
\begin{equation}
    \int_{\Sigma_g}\rho_I\wedge\sigma_J=\delta_{IJ},
    \qquad
    \int_{\Sigma_g}\rho_I\wedge\rho_J
    =
    \int_{\Sigma_g}\sigma_I\wedge\sigma_J=0 .
    \label{eq:higher-genus-intersection-basis}
\end{equation}
The flat background gauge field and the localized dynamical gauge field are
expanded as
\begin{equation}
    A_{\rm flat}^a
    =
    \sum_{I=1}^g(\theta_I^a\rho_I+\nu_I^a\sigma_I),
    \qquad
    B_{\rm flat}^a
    =
    \sum_{I=1}^g
    (\widetilde\theta_I^a\rho_I+\widetilde\nu_I^a\sigma_I),
    \qquad
    a=1,2 .
    \label{eq:higher-genus-flat-holonomies}
\end{equation}
Large gauge transformations shift each lifted holonomy by an integral multiple
of \(2\pi\).  In the deformed partition function below the dynamical holonomies
are integrated over their lifted real values, not over a fundamental domain.

For two harmonic one-forms
\(\alpha=\sum_I(\theta_{\alpha,I}\rho_I+\nu_{\alpha,I}\sigma_I)\) and
\(\beta=\sum_I(\theta_{\beta,I}\rho_I+\nu_{\beta,I}\sigma_I)\),
\begin{equation}
    \int_{\Sigma_g}\alpha\wedge\beta
    =
    \sum_{I=1}^g
    \left(
    \theta_{\alpha,I}\nu_{\beta,I}
    -
    \nu_{\alpha,I}\theta_{\beta,I}
    \right).
    \label{eq:higher-genus-harmonic-pairing}
\end{equation}

Define
\begin{equation}
    \Delta\theta_I^a=\theta_I^a-\widetilde\theta_I^a,
    \qquad
    \Delta\nu_I^a=\nu_I^a-\widetilde\nu_I^a .
    \label{eq:higher-genus-holonomy-differences}
\end{equation}
With this Darboux basis for \(H^1(\Sigma_g,\mathbb R)\), the harmonic
finite-dimensional kernel is
\begin{equation}
\begin{aligned}
    \mathcal K_{\lambda,C}^{(g)}
    (\theta_I^a,\nu_I^a;\widetilde\theta_I^a,\widetilde\nu_I^a)
    &=
    \exp\!\Bigg\{\sum_{I=1}^g\bigg[
    C_{ab}(\widetilde\theta_I^a\Delta\nu_I^b-\widetilde\nu_I^a\Delta\theta_I^b)
    \\
    &\hspace{28mm}
    +
    \left(\frac12 C_{ab}-\frac{i}{4\lambda}\epsilon_{ab}\right)
    (\Delta\theta_I^a\Delta\nu_I^b-\Delta\nu_I^a\Delta\theta_I^b)
    \bigg]\Bigg\}.
\end{aligned}
    \label{eq:higher-genus-flat-kernel}
\end{equation}
Equivalently, it is obtained from the torus finite-dimensional kernel
\eqref{eq:torus-flat-kernel} by replacing the genus-one pairing with
\eqref{eq:higher-genus-harmonic-pairing}.

With the background-independent normalization fixed by the undeformed limit, the
deformed genus-\(g\) partition function with flat background gauge fields is
\begin{equation}
\begin{aligned}
    Z_{\lambda,C}^{(g)}[A_{\rm flat}]
    =
    \left[
    \frac{|\det\mathsf M_{\lambda,C}|}{(2\pi)^2}
    \right]^g
    \prod_{a=1}^{2}\prod_{I=1}^{g}
    \int_{-\infty}^{\infty}
    d\widetilde\theta_I^a\,d\widetilde\nu_I^a\,
    Z_0^{(g)}[B_{\rm flat}]\,
    \mathcal K_{\lambda,C}^{(g)}
    (\theta_I^a,\nu_I^a;\widetilde\theta_I^a,\widetilde\nu_I^a) .
\end{aligned}
    \label{eq:higher-genus-flat-kernel-transform}
\end{equation}
The power \(g\) reflects the product over the \(g\) independent harmonic pairs.  Since
\(\det\mathsf M_{\lambda,C}=1/(4\lambda^2)+O(1)\), the oscillatory part of the
kernel gives
\begin{equation}
    \left[
    \frac{|\det\mathsf M_{\lambda,C}|}{(2\pi)^2}
    \right]^g
    \mathcal K_{\lambda,C}^{(g)}
    \longrightarrow
    \prod_{a=1}^{2}\prod_{I=1}^{g}
    \delta(\theta_I^a-\widetilde\theta_I^a)\,
    \delta(\nu_I^a-\widetilde\nu_I^a).
    \label{eq:higher-genus-undeformed-limit}
\end{equation}
So \eqref{eq:higher-genus-flat-kernel-transform} reduces to
\(Z_0^{(g)}[A_{\rm flat}]\).  Background-independent determinants
from the non-zero modes, together with possible curvature counterterms depending
only on the topology of \(\Sigma_g\), are part of the overall normalization
scheme.  They do not affect the flat-background dependence or the finite gauge
anomaly.

\section{Examples}
\label{sec:examples}

The examples in this section test the flat-background kernel in complementary
settings.  The compact boson is an anomalous conformal example: at zero
background the deformation shifts the dynamical radius, while in flat
backgrounds the anomaly scheme continues to control the contact terms and
spectral-flow data.  The massive examples are anomaly-free and non-conformal.  The complex bosons lead to a non-polynomial local interaction, while the Dirac fermions give the
Federbush model, whose rapidity-independent scattering phase appears as a
charge-dependent twist shift in finite volume.

\subsection{The compact boson and Abelian WZW model}
\label{sec:compact-free-boson-example}
We now test the anomalous double-current transform in the compact free boson,
equivalently the Abelian \(U(1)\) WZW model.  The model is
elementary, but it captures the main subtlety of the anomalous construction:
the deformation changes the dynamical radius parameter, while the anomaly
scheme fixes the contact terms and spectral-flow data.  We first choose an
explicit anomalous coupling to background \(U(1)_L\times U(1)_R\) gauge fields,
then compute the local deformed action and the exact torus flat-background
partition function.  The final subsection records the corresponding
same-chirality and opposite-chirality deformations for two compact bosons.

\subsection*{The undeformed theory and its anomaly}
\label{subsec:free-boson-undeformed-anomalous-backgrounds}

The seed theory is a compact boson \(\varphi\sim\varphi+2\pi\), equivalently
the Abelian \(U(1)\) WZW model \cite{Witten:1991mm}.  We couple its two chiral
\(U(1)\) symmetries to background gauge fields \(A^L\) and \(A^R\), written in
complex coordinates as
\begin{equation}
    A^a=A_z^a dz+A_{\bar z}^a d\bar z .
    \label{eq:free-boson-background-components}
\end{equation}
The action at vanishing background gauge field is
\begin{equation}
    S_{U(1)}[\varphi]
    =
    \frac{k}{4\pi}
    \int_\Sigma d^2z\,\partial\varphi\,\bar\partial\varphi ,
    \label{eq:free-boson-action}
\end{equation}
where \(k>0\) is the current algebra level in this normalization.  We choose the
following anomalous background-field coupling:
\begin{equation}
\begin{aligned}
    S_{U(1)}^{\rm gauged}[\varphi;A^L,A^R]
    &=
    \frac{k}{4\pi}
    \int_\Sigma d^2z\,\partial\varphi\,\bar\partial\varphi
    \\
    &\quad+
    \frac{k}{2\pi}
    \int_\Sigma d^2z\,
    \left[
    -A_{\bar z}^R\,\partial\varphi
    +A_z^L\,\bar\partial\varphi
    -A_z^LA_{\bar z}^R
    +\frac12 A_z^RA_{\bar z}^R
    +\frac12 A_z^LA_{\bar z}^L
    \right].
\end{aligned}
    \label{eq:free-boson-gauged-action}
\end{equation}
The undeformed generating functional in this scheme is
\begin{equation}
    Z_{U(1)}[A^L,A^R]
    =
    \int[D\varphi]\,
    \exp\!\left[-S_{U(1)}^{\rm gauged}[\varphi;A^L,A^R]\right].
    \label{eq:free-boson-generating-functional}
\end{equation}

For a \(U(1)\)-valued function \(U\), set
\begin{equation}
    \vartheta(U)=-i\,U^{-1}dU .
    \label{eq:free-boson-u1-one-form}
\end{equation}
Thus \(\vartheta(U)=d\chi\) locally when \(U=e^{i\chi}\); large transformations
are included by allowing non-zero periods of \(\vartheta(U)\).  The finite
background transformation is
\begin{equation}
    e^{i\varphi}\mapsto U^Le^{i\varphi}(U^R)^{-1},
    \qquad
    A^a\mapsto A^a-\vartheta(U^a).
    \label{eq:free-boson-background-gauge-transform}
\end{equation}
The coupling \eqref{eq:free-boson-gauged-action} realizes the \(C\)-anomaly
representative of Section \ref{subsec:anomalous-transform}.  In the local
trivialization used here, the finite transformation law is
\begin{equation}
    Z_{U(1)}[A^L-\vartheta(U^L),A^R-\vartheta(U^R)]
    =
    \exp\!\left[-\mathcal I^{U(1)}_{\rm anom}[U,A]\right]
    Z_{U(1)}[A^L,A^R],
    \label{eq:free-boson-finite-anomaly-law}
\end{equation}
where
\begin{equation}
    \mathcal I^{U(1)}_{\rm anom}[U,A]
    =
    -\frac{ik}{4\pi}\int_\Sigma\vartheta(U^L)\wedge A^L
    +
    \frac{ik}{4\pi}\int_\Sigma\vartheta(U^R)\wedge A^R .
    \label{eq:free-boson-finite-anomaly}
\end{equation}
Equivalently, in the ordered basis \(a=(L,R)\), the anomaly matrix is
\begin{equation}
    C^{U(1)}_{ab}
    =
    \frac{ik}{4\pi}
    \begin{pmatrix}
        -1 & 0 \\
        0 & 1
    \end{pmatrix}_{ab}.
    \label{eq:free-boson-anomaly-matrix}
\end{equation}
For \(U^a=e^{i\varepsilon^a}\), this gives the infinitesimal anomaly
\begin{equation}
    \mathcal I^{U(1)}_{\rm anom}[e^{i\varepsilon},A]
    =
    -\frac{ik}{4\pi}
    \int_\Sigma
    \left(\varepsilon^R\,dA^R-\varepsilon^L\,dA^L\right)
    +O(\varepsilon^2).
    \label{eq:free-boson-infinitesimal-anomaly}
\end{equation}

We define background-field currents by varying these background gauge fields,
with chiral signs chosen so that the currents at vanishing background have the
standard Kac-Moody normalization:
\begin{equation}
    \begin{aligned}
    \delta S_{U(1)}^{\rm gauged}
    =
    -\frac{i}{2\pi}\int_\Sigma J^L\wedge\delta A^L
    +
    \frac{i}{2\pi}\int_\Sigma J^R\wedge\delta A^R
    +
    \delta_\varphi S .
    \end{aligned}
    \label{eq:free-boson-current-definition}
\end{equation}
Here \(\delta_\varphi S\) denotes terms proportional to the
\(\varphi\)-equation of motion, which is
\begin{equation}
    \partial\bar\partial\varphi
    -\partial A_{\bar z}^R
    +\bar\partial A_z^L
    =
    0 .
    \label{eq:free-boson-eom}
\end{equation}
The resulting current components are
\begin{equation}
\begin{aligned}
    J_{\bar z}^L
    &=
    k\left(\bar\partial\varphi-A_{\bar z}^R+\frac12 A_{\bar z}^L\right),
    &
    J_z^L
    &=
    -\frac{k}{2}A_z^L,
    \\
    J_z^R
    &=
    -k\left(\partial\varphi+A_z^L-\frac12 A_z^R\right),
    &
    J_{\bar z}^R
    &=
    -\frac{k}{2}A_{\bar z}^R .
\end{aligned}
    \label{eq:free-boson-current-components}
\end{equation}
Writing \(F^a_{z\bar z}=\partial A_{\bar z}^a-\bar\partial A_z^a\), the
equation of motion implies the anomalous Ward identities
\begin{equation}
    \partial J_{\bar z}^L-\bar\partial J_z^L
    =
    \frac{k}{2}F^L_{z\bar z},
    \qquad
    \partial J_{\bar z}^R-\bar\partial J_z^R
    =
    \frac{k}{2}F^R_{z\bar z}.
    \label{eq:free-boson-current-ward-identities}
\end{equation}
Thus the background-field currents obey the expected anomalous conservation law
and are conserved in flat backgrounds.  At vanishing background,
\begin{equation}
    J_{\bar z}^L=k\bar\partial\varphi,
    \qquad
    J_z^R=-k\partial\varphi,
    \qquad
    \partial J_{\bar z}^L=\bar\partial J_z^R=0 .
    \label{eq:free-boson-vanishing-background-currents}
\end{equation}
These are the two chiral currents to which the anomalous double-current
deformation of
Section \ref{subsec:anomalous-transform} is applied.

\subsection*{Classical deformed action}
\label{subsec:free-boson-classical-deformed-action}

We first isolate the local classical effect of the deformation.  This is the
local saddle of the path integral, before restoring the
global holonomy and winding sectors which enter the torus partition function.
Specializing \eqref{eq:anomalous-transform} to
\eqref{eq:free-boson-anomaly-matrix}, the dynamical fields are a pair of
connections \(B^a\) and compact Stueckelberg fields \(h^a\), entering through
\begin{equation}
    \Theta^a=A^a-B^a+\vartheta(h^a).
    \label{eq:free-boson-classical-Theta}
\end{equation}
For the local saddle one can use the dynamical gauge redundancy to impose the
unitary gauge
\begin{equation}
    h^L=h^R=1,
    \qquad
    \Theta^a=A^a-B^a .
    \label{eq:free-boson-classical-unitary-gauge}
\end{equation}
This gauge choice is only local: windings of \(h^a\) and flat holonomies of
\(B^a\) will be restored in the exact path integral.  In
\eqref{eq:free-boson-classical-unitary-gauge}, however, \(B^a\) enters the
local Lagrangian algebraically, so its elimination is a Gaussian saddle problem.
Substituting the saddle back into the auxiliary action gives
\begin{equation}
\begin{aligned}
    S_{U(1),\lambda}^{\rm cl}[\varphi;A^L,A^R]
    &=
    \left.
    S_{U(1)}^{\rm gauged}[\varphi;A^L,A^R]
    \right|_{k\to K_\lambda}
    \\
    &\quad
    +
    \frac{K_\lambda-k}{4\pi}
    \int_\Sigma d^2z\,
    \left[
    \partial\varphi\,(A_{\bar z}^L+A_{\bar z}^R)
    -(A_z^L+A_z^R)\,\bar\partial\varphi
    +A_z^LA_{\bar z}^R-A_z^RA_{\bar z}^L
    \right].
\end{aligned}
    \label{eq:free-boson-classical-action}
\end{equation}
The notation in the first term means \eqref{eq:free-boson-gauged-action} with
\(k\) replaced by \(K_\lambda\).  The second line is the extra
background-dependent term required by the fixed anomaly scheme.  In
particular, at vanishing background one obtains again a compact-boson action,
\begin{equation}
    S_{U(1),\lambda}^{\rm cl}[\varphi;0,0]
    =
    \frac{K_\lambda}{4\pi}
    \int_\Sigma d^2z\,\partial\varphi\,\bar\partial\varphi,
    \qquad
    K_\lambda
    =
    k\left(\frac{1+\frac{k\lambda}{2\pi}}
    {1-\frac{k\lambda}{2\pi}}\right)^2 .
    \label{eq:free-boson-classical-Klambda}
\end{equation}
Thus the local dynamics at vanishing background is the radius shift
\(k\to K_\lambda\).
The background-field-coupled action, however, is not obtained by making this
replacement in \eqref{eq:free-boson-gauged-action}: the  contact term is
still tied to the original anomaly coefficient \(k\), and the second line of
\eqref{eq:free-boson-classical-action} supplies the compensating
background-dependent interaction.

The same background-current convention as
\eqref{eq:free-boson-current-definition} defines the classical deformed
currents.  Varying \eqref{eq:free-boson-classical-action} and then setting
\(A^L=A^R=0\) gives
\begin{equation}
\begin{aligned}
    (J_\lambda^L)_{\bar z}
    &=
    \frac{K_\lambda+k}{2}\,\bar\partial\varphi,
    &
    (J_\lambda^L)_z
    &=
    -\frac{K_\lambda-k}{2}\,\partial\varphi,
    \\
    (J_\lambda^R)_z
    &=
    -\frac{K_\lambda+k}{2}\,\partial\varphi,
    &
    (J_\lambda^R)_{\bar z}
    &=
    \frac{K_\lambda-k}{2}\,\bar\partial\varphi .
\end{aligned}
    \label{eq:free-boson-vanishing-background-deformed-currents}
\end{equation}
They are conserved on the equation of motion at vanishing background,
\(\partial(J_\lambda^a)_{\bar z}-\bar\partial(J_\lambda^a)_z=0\), but for
finite \(\lambda\) they are not purely chiral.

For comparison with the torus trace, we identify the zero modes in a spatially
twisted sector.  Using cylinder coordinates \((t,\sigma)\), with
\(\sigma=\sigma^1\), the time components of
\eqref{eq:free-boson-vanishing-background-deformed-currents} are
\begin{equation}
    J^L_{\lambda,t}
    =
    -\frac{K_\lambda}{2}\partial_\sigma\varphi
    +\frac{k}{2}\partial_t\varphi,
    \qquad
    J^R_{\lambda,t}
    =
    -\frac{K_\lambda}{2}\partial_\sigma\varphi
    -\frac{k}{2}\partial_t\varphi .
    \label{eq:free-boson-cylinder-current-components}
\end{equation}
Turn on a spatial flat background
\(A^a_\sigma=\theta^a/(2\pi)\), \(A^a_t=0\).  The background coupling linear in
\(A_\sigma^a\) is
\begin{equation}
\begin{aligned}
    \mathcal L_{\rm lin}
    &=
    \frac{1}{2\pi}
    \left(A^L_\sigma J^L_{\lambda,t}
    -
    A^R_\sigma J^R_{\lambda,t}\right)
    \\
    &=
    -\frac{K_\lambda}{4\pi}(A^L_\sigma-A^R_\sigma)\partial_\sigma\varphi
    +
    \frac{k}{4\pi}(A^L_\sigma+A^R_\sigma)\partial_t\varphi .
\end{aligned}
    \label{eq:free-boson-spatial-background-linear-coupling}
\end{equation}
The first term combines with the spatial kinetic term into
\(-K_\lambda(\partial_\sigma\varphi+A^L_\sigma-A^R_\sigma)^2/(8\pi)\), up to a
local background-field term.  Hence
\(\oint d\sigma\,(\partial_\sigma\varphi+A^L_\sigma-A^R_\sigma)
=2\pi w_\theta\), with
\(w_\theta=w+(\theta^L-\theta^R)/(2\pi)\).  The second term shifts the
canonical momentum:
\begin{equation}
    \Pi_\varphi
    =
    \frac{K_\lambda}{4\pi}\partial_t\varphi
    +
    \frac{k}{4\pi}(A^L_\sigma+A^R_\sigma),
    \qquad
    \ell_\theta
    =
    \frac{K_\lambda}{4\pi}\oint d\sigma\,\partial_t\varphi
    =
    \ell-\frac{k}{4\pi}(\theta^L+\theta^R).
    \label{eq:free-boson-twisted-momentum-zero-mode}
\end{equation}
The left- and right-moving zero modes therefore obey
\begin{equation}
    \frac{1}{2\pi}\int d\sigma^1\,K_\lambda\partial\varphi
    =
    \ell_\theta+\frac{K_\lambda}{2}w_\theta,
    \qquad
    \frac{1}{2\pi}\int d\sigma^1\,K_\lambda\bar\partial\varphi
    =
    -\ell_\theta+\frac{K_\lambda}{2}w_\theta .
    \label{eq:free-boson-classical-zero-mode-integrals}
\end{equation}
Therefore \eqref{eq:free-boson-vanishing-background-deformed-currents} gives
\begin{equation}
\begin{aligned}
    Q_{L,\lambda,\theta}
    &=
    -\ell_\theta+\frac{k}{2}w_\theta
    =
    -\ell+\frac{k}{2}w+\frac{k}{2\pi}\theta^L,
    \\
    Q_{R,\lambda,\theta}
    &=
    \ell_\theta+\frac{k}{2}w_\theta
    =
    \ell+\frac{k}{2}w-\frac{k}{2\pi}\theta^R .
\end{aligned}
    \label{eq:free-boson-classical-deformed-charges}
\end{equation}
Thus the local current contains \(K_\lambda\), while the integrated charges in
the fixed anomaly scheme are controlled by the original level \(k\).  The exact
flat-background transform below reproduces the same charge assignments.

\subsection*{Deformed torus partition function}
\label{subsec:free-boson-deformed-torus-partition-function}

We now evaluate the genus-one flat-background transform.  The background gauge
field is a flat representative of \(A^a\), written as in
\eqref{eq:torus-flat-background} and denoted here by \(A_{\rm flat}^a\).  The
dynamical flat gauge field is
\begin{equation}
    B_{\rm flat}^a
    =
    \widetilde\theta^a\rho+\widetilde\nu^a\sigma,
    \qquad
    \widetilde\theta^a,\widetilde\nu^a\in\mathbb R,
    \qquad
    a=L,R .
    \label{eq:free-boson-flat-dynamical-field}
\end{equation}
We use the abbreviations
\begin{equation}
    p_{m,n}=n-m\tau,
    \qquad
    \bar p_{m,n}=n-m\bar\tau,
    \qquad
    \widetilde u^a=\widetilde\nu^a-\tau\widetilde\theta^a,
    \qquad
    \widetilde{\bar{u}}^a=\widetilde\nu^a-\bar\tau\widetilde\theta^a .
    \label{eq:free-boson-holonomy-shorthand}
\end{equation}
With the compact-boson normalization fixed as in Appendix \ref{app:flat-background-path integral-details}, the undeformed
flat-background partition function is the standard winding-sum representation
of the compact boson torus partition function
\cite{Ginsparg:1988ui,Polchinski:1998rq}, written here in the
background-field conventions of \cite{Witten:1991mm,Murthy:2025ioh},
\begin{equation}
    Z_{U(1)}[B_{\rm flat}^L,B_{\rm flat}^R]
    =
    \sqrt{\frac{k}{2\tau_2}}\,
    \frac{1}{|\eta(\tau)|^2}
    \sum_{m,n\in\mathbb Z}
    \exp[-S^k_{m,n}(\widetilde\theta^a,\widetilde\nu^a)] ,
    \label{eq:free-boson-undeformed-winding-sum}
\end{equation}
where
\begin{equation}
\begin{aligned}
    S^k_{m,n}
    &=
    S^k_{\rm loc}(\widetilde\theta^a,\widetilde\nu^a)
    +
    \frac{\pi k}{2\tau_2}\,p_{m,n}\bar p_{m,n}
    +
    \frac{k}{2\tau_2}
    \left(\widetilde{\bar{u}}^Lp_{m,n}-\widetilde u^R\bar p_{m,n}\right),
    \\
    S^k_{\rm loc}
    &=
    \frac{k}{8\pi\tau_2}
    \left[
    \widetilde{\bar{u}}^L\widetilde u^L
    +\widetilde{\bar{u}}^R\widetilde u^R
    -2\widetilde{\bar{u}}^L\widetilde u^R
    \right].
\end{aligned}
    \label{eq:free-boson-undeformed-winding-action}
\end{equation}
The factor \(|\eta(\tau)|^{-2}\) is the oscillator denominator, and
\(\sqrt{k/(2\tau_2)}\) is the zero-mode normalization in this winding
representation.  We also record the corresponding Hilbert-space form.  After
Poisson resummation of the temporal winding, a state is labelled by
\((w,\ell,N_L,N_R)\), with \(w,\ell\in\mathbb Z\) and \(N_L,N_R\geq0\).  For
spatial holonomies \(\theta^a\), define
\begin{equation}
    w_\theta=w+\frac{\theta^L-\theta^R}{2\pi},
    \qquad
    \ell_\theta=\ell-\frac{k}{4\pi}\left(\theta^L+\theta^R\right).
    \label{eq:free-boson-shifted-zero-mode-variables}
\end{equation}
The charges which couple to the temporal holonomies are
\begin{equation}
\begin{aligned}
    Q_{L,\theta}(\theta,w,\ell)
    &=
    -\ell_\theta+\frac{k}{2}w_\theta
    =
    -\ell+\frac{k}{2}w+\frac{k}{2\pi}\theta^L,
    \\
    Q_{R,\theta}(\theta,w,\ell)
    &=
    \ell_\theta+\frac{k}{2}w_\theta
    =
    \ell+\frac{k}{2}w-\frac{k}{2\pi}\theta^R .
\end{aligned}
    \label{eq:free-boson-charges-explicit}
\end{equation}
The Hamiltonian and momentum in the \(\theta\)-twisted sector are
\begin{equation}
\begin{aligned}
    H_\theta(\theta,w,\ell,N_L,N_R)
    &=
    \frac{1}{k}
    \left[\ell-\frac{k}{4\pi}\left(\theta^L+\theta^R\right)\right]^2
    +
    \frac{k}{4}
    \left[w+\frac{\theta^L-\theta^R}{2\pi}\right]^2
    +
    N_L+N_R-\frac{1}{12},
    \\
    P_\theta(\theta,w,\ell,N_L,N_R)
    &=
    \left[\ell-\frac{k}{4\pi}\left(\theta^L+\theta^R\right)\right]
    \left[w+\frac{\theta^L-\theta^R}{2\pi}\right]
    +
    N_L-N_R .
\end{aligned}
    \label{eq:free-boson-energy-momentum-explicit}
\end{equation}
For the dynamical flat gauge field \(B_{\rm flat}\), the same formulas hold with
\(\theta^a\) replaced by \(\widetilde\theta^a\).  The  contact term in
the anomaly scheme of \eqref{eq:free-boson-finite-anomaly} is
\begin{equation}
    S_{\rm contact}(\theta^a,\nu^a)
    =
    \frac{ik}{4\pi}
    \left(
    \nu^L\theta^L
    -
    \nu^R\theta^R
    \right).
    \label{eq:free-boson-undeformed-contact-term}
\end{equation}
Thus, for the background flat gauge field,
\begin{equation}
\begin{aligned}
    Z_{U(1)}[A_{\rm flat}^L,A_{\rm flat}^R]
    &=
    \exp[-S_{\rm contact}(\theta^a,\nu^a)]
    \\
    &\quad\times
    {\rm Tr}_{\mathcal H_{\theta^L,\theta^R}}
    \exp\!\left[
    -2\pi\tau_2 H_\theta
    +2\pi i\tau_1 P_\theta
    +i\nu^L Q_{L,\theta}
    +i\nu^R Q_{R,\theta}
    \right],
\end{aligned}
    \label{eq:free-boson-undeformed-flat-trace}
\end{equation}
where the eigenvalues of \(H_\theta\), \(P_\theta\), \(Q_{L,\theta}\), and
\(Q_{R,\theta}\) are those in \eqref{eq:free-boson-charges-explicit} and
\eqref{eq:free-boson-energy-momentum-explicit}.  The partition function
evaluated at \(B_{\rm flat}\) is obtained from the same expression by replacing
\((\theta^a,\nu^a)\) with \((\widetilde\theta^a,\widetilde\nu^a)\).

In this anomaly scheme, the genus-one finite-dimensional kernel
\eqref{eq:torus-flat-kernel} becomes
\begin{equation}
\begin{aligned}
    \mathcal K_\lambda^{U(1)}
    =
    \exp\!\Bigg[
    -\frac{i}{2\lambda}
    \left(\Delta\theta^L\Delta\nu^R-\Delta\nu^L\Delta\theta^R\right)
    +\frac{ik}{4\pi}
    \left(
    -\widetilde\theta^L\nu^L+\widetilde\nu^L\theta^L
    +\widetilde\theta^R\nu^R-\widetilde\nu^R\theta^R
    \right)
    \Bigg],
\end{aligned}
    \label{eq:free-boson-specialized-kernel}
\end{equation}
with \(\Delta\theta^a=\theta^a-\widetilde\theta^a\) and
\(\Delta\nu^a=\nu^a-\widetilde\nu^a\).  Moreover
\begin{equation}
    \det\mathsf M_{\lambda,C^{U(1)}}
    =
    \frac{1}{4\lambda^2}
    -
    \frac{k^2}{16\pi^2}.
    \label{eq:free-boson-M-determinant}
\end{equation}
For non-zero \(\lambda\), and away from the degeneracy locus
\(\lambda=\pm2\pi/k\), the deformed flat-background partition function is
\begin{equation}
\begin{aligned}
    Z_{U(1),\lambda}[A_{\rm flat}^L,A_{\rm flat}^R]
    =
    \frac{1}{16\pi^2\lambda^2}
    \left|1-\frac{k^2\lambda^2}{4\pi^2}\right|
    \prod_{a=L,R}\int_{-\infty}^{\infty}
    d\widetilde\theta^a\,d\widetilde\nu^a
    Z_{U(1)}[B_{\rm flat}^L,B_{\rm flat}^R]\,
    \mathcal K_\lambda^{U(1)} .
\end{aligned}
    \label{eq:free-boson-flat-kernel-transform}
\end{equation}
This is the flat-background transform \eqref{eq:torus-flat-kernel-transform}
specialized to the free-boson anomaly scheme, with the flat-measure
normalization derived in Appendix
\ref{app:flat-background-path integral-details}.

The remaining integral over dynamical holonomies is Gaussian in each winding
sector.  For the background holonomies set
\begin{equation}
    u^a=\nu^a-\tau\theta^a,
    \qquad
    \bar u^a=\nu^a-\bar\tau\theta^a .
    \label{eq:free-boson-background-holonomy-shorthand}
\end{equation}
Using the same \(K_\lambda\) as in
\eqref{eq:free-boson-classical-Klambda}, one obtains
\begin{equation}
    \frac{1}{16\pi^2\lambda^2}
    \left|1-\frac{k^2\lambda^2}{4\pi^2}\right|
    \prod_{a=L,R}\int d\widetilde\theta^a\,d\widetilde\nu^a\,
    \mathcal K_\lambda^{U(1)}e^{-S^k_{m,n}(\widetilde\theta,\widetilde\nu)}
    =
    \sqrt{\frac{K_\lambda}{k}}\,
    e^{-S^{(\lambda)}_{m,n}(\theta,\nu)} .
    \label{eq:free-boson-holonomy-gaussian}
\end{equation}
The deformed winding action is
\begin{equation}
    S^{(\lambda)}_{m,n}
    =
    S_{m,n}^{K_\lambda}
    +
    \frac{K_\lambda-k}{4\tau_2}
    \left[
    (u^L+u^R)\bar p_{m,n}
    -
    (\bar u^L+\bar u^R)p_{m,n}
    +
    \frac{\bar u^Lu^R-\bar u^Ru^L}{2\pi}
    \right].
    \label{eq:free-boson-deformed-winding-action}
\end{equation}
Here \(S_{m,n}^{K_\lambda}\) is \eqref{eq:free-boson-undeformed-winding-action}
with \(k\) replaced by \(K_\lambda\) and with the dynamical holonomies replaced
by the background holonomies.  Combining
\eqref{eq:free-boson-undeformed-winding-sum} with
\eqref{eq:free-boson-holonomy-gaussian} gives
\begin{equation}
    Z_{U(1),\lambda}[A_{\rm flat}^L,A_{\rm flat}^R]
    =
    \sqrt{\frac{K_\lambda}{2\tau_2}}\,
    \frac{1}{|\eta(\tau)|^2}
    \sum_{m,n\in\mathbb Z}
    \exp[-S^{(\lambda)}_{m,n}(\theta^a,\nu^a)] .
    \label{eq:free-boson-deformed-winding-sum}
\end{equation}
Equation \eqref{eq:free-boson-deformed-winding-sum} is the direct output of the
path integral transform.  Reading it as a spectrum requires a convention for
separating local background-field terms from background-dependent spectral flow.
We keep the finite anomaly representative \eqref{eq:free-boson-finite-anomaly},
and hence the  contact term
\eqref{eq:free-boson-undeformed-contact-term}, fixed.
In this convention, Poisson resummation of the temporal winding, with \(m=w\),
gives
\begin{equation}
\begin{aligned}
    Z_{U(1),\lambda}[A_{\rm flat}^L,A_{\rm flat}^R]
    &=
    \exp[-S_{\rm contact}(\theta^a,\nu^a)]
    \\
    &\quad\times
    {\rm Tr}_{\mathcal H_{\theta^L,\theta^R}}
    \exp\!\left[
    -2\pi\tau_2H_{\lambda,\theta}
    +2\pi i\tau_1P_{\lambda,\theta}
    +i\nu^LQ_{L,\lambda,\theta}
    +i\nu^RQ_{R,\lambda,\theta}
    \right].
\end{aligned}
    \label{eq:free-boson-deformed-flat-trace}
\end{equation}
The eigenvalues on a state labelled by \((w,\ell,N_L,N_R)\) are
\begin{equation}
\begin{aligned}
    H_{\lambda,\theta}(\theta,w,\ell,N_L,N_R)
    &=
    \frac{\ell_\theta^2}{K_\lambda}
    +
    \frac{K_\lambda}{4}w_\theta^2
    +
    N_L+N_R-\frac{1}{12},
    \\
    P_{\lambda,\theta}(\theta,w,\ell,N_L,N_R)
    &=
    \ell_\theta w_\theta
    +
    N_L-N_R,
\end{aligned}
    \label{eq:free-boson-deformed-energy-momentum}
\end{equation}
while the charge eigenvalues are
\begin{equation}
\begin{aligned}
    Q_{L,\lambda,\theta}(\theta,w,\ell)
    &=
    -\ell_\theta+\frac{k}{2}w_\theta
    =
    -\ell+\frac{k}{2}w+\frac{k}{2\pi}\theta^L,
    \\
    Q_{R,\lambda,\theta}(\theta,w,\ell)
    &=
    \ell_\theta+\frac{k}{2}w_\theta
    =
    \ell+\frac{k}{2}w-\frac{k}{2\pi}\theta^R .
\end{aligned}
    \label{eq:free-boson-deformed-charges}
\end{equation}
The relation to the undeformed twisted spectrum is transparent after introducing
\begin{equation}
    e^{2\gamma_\lambda}
    =
    \frac{K_\lambda}{k},
    \qquad
    \gamma_\lambda
    =
    \log\!\left(
    \frac{1+\frac{k\lambda}{2\pi}}
    {1-\frac{k\lambda}{2\pi}}
    \right).
    \label{eq:free-boson-boost-parameter}
\end{equation}
The charges in \eqref{eq:free-boson-deformed-charges} coincide with the
undeformed charges \eqref{eq:free-boson-charges-explicit}.  Therefore the
Hamiltonian deformation can be written entirely in terms of undeformed spectral
data:
\begin{equation}
    H_{\lambda,\theta}
    =
    H_\theta
    +
    \frac{\sinh(2\gamma_\lambda)}{k}\,
    Q_{L,\theta}Q_{R,\theta}
    +
    \frac{\sinh^2\gamma_\lambda}{k}
    \left(Q_{L,\theta}^2+Q_{R,\theta}^2\right),
    \qquad
    P_{\lambda,\theta}=P_\theta .
    \label{eq:free-boson-deformed-undeformed-relation}
\end{equation}
These charges agree with the classical zero-mode result
\eqref{eq:free-boson-classical-deformed-charges}.  The deformation therefore changes
the kinetic coefficient in the spectrum at vanishing background from \(k\) to
\(K_\lambda\),
while the spectral-flow charges and  contact
term remain controlled by the original anomaly coefficient \(k\).  The
background-field-coupled partition function is not obtained by replacing \(k\) by
\(K_\lambda\) everywhere.

The split between contact terms and background-dependent spectral data is a scheme
choice.  The fixed-scheme statement above is a convenient way to present the
trace, but the conclusion at vanishing background is independent of this split.
Setting \(\theta^a=\nu^a=0\), the extra background-dependent terms in
\eqref{eq:free-boson-deformed-winding-action} vanish and
\begin{equation}
    Z_{U(1),\lambda}[0,0]
    =
    \sqrt{\frac{K_\lambda}{2\tau_2}}\,
    \frac{1}{|\eta(\tau)|^2}
    \sum_{m,n\in\mathbb Z}
    \exp[-S_{m,n}^{K_\lambda}(0,0)] .
    \label{eq:free-boson-vanishing-background-deformed-winding-sum}
\end{equation}
Thus, without background holonomies, the deformed theory has the spectrum of the
compact free boson with radius parameter \(K_\lambda\).  Finally, as
\(\lambda\to0\), one has \(K_\lambda\to k\) and the background-dependent correction
in \eqref{eq:free-boson-deformed-winding-action} vanishes.  Hence
\eqref{eq:free-boson-deformed-winding-sum} reduces to the undeformed result
\eqref{eq:free-boson-undeformed-winding-sum} evaluated at the background flat
gauge field.

\subsection*{Double-current deformations for two compact bosons}
\label{subsec:two-abelian-chirality-checks}

Now we consider the double-current deformations for two compact bosons \(\phi_I\sim\phi_I+2\pi\),
\(I=1,2\), with levels \(k_I>0\).  They realize the standard distinction
between same-chirality and opposite-chirality double-current deformations
\cite{Borsato:2023dis}.  Only the choice of chiral directions is changed.  The
symbols \(S_0^{R_1R_2}\) and \(S_0^{L_1R_2}\) denote the corresponding
undeformed gauged actions.

\paragraph{$U(1)^R_1\times U(1)^R_2$ case:}

Choose both directions to be right-moving,
\begin{equation}
    (a=1,a=2)=(R_1,R_2),
    \qquad
    C^{R_1R_2}_{ab}
    =
    \frac{i}{4\pi}
    \begin{pmatrix}
        k_1&0\\
        0&k_2
    \end{pmatrix}_{ab}.
    \label{eq:RR-embedding}
\end{equation}
Both selected currents are of type \((1,0)\) at vanishing background.  The
density \(J^{R_1}\wedge J^{R_2}\) therefore vanishes, and the
local deformation is trivial:
\begin{equation}
    S^{R_1R_2}_{\lambda,{\rm cl}}[\phi_1,\phi_2;A^1,A^2]
    =
    S^{R_1R_2}_{0}[\phi_1,\phi_2;A^1,A^2].
    \label{eq:RR-trivial-classical-action}
\end{equation}
The flat-background partition function is unchanged for the same reason:
\begin{equation}
    Z^{R_1R_2}_{\lambda}[A^1_{\rm flat},A^2_{\rm flat}]
    =
    Z^{R_1R_2}_{0}[A^1_{\rm flat},A^2_{\rm flat}] .
    \label{eq:RR-trivial-flat-background-partition-function}
\end{equation}

\paragraph{$U(1)^L_1\times U(1)^R_2$ case:}

Choose instead opposite chiralities,
\begin{equation}
    (a=1,a=2)=(L_1,R_2),
    \qquad
    C^{L_1R_2}_{ab}
    =
    \frac{i}{4\pi}
    \begin{pmatrix}
        -k_1&0\\
        0&k_2
    \end{pmatrix}_{ab}.
    \label{eq:LR-embedding-two-bosons}
\end{equation}
The saddle now produces a non-trivial deformation.  At vanishing background one
finds
\begin{equation}
    S^{L_1R_2}_{\lambda,{\rm cl}}[\phi_1,\phi_2;0,0]
    =
    \frac{1}{4\pi}\int d^2z\,
    \left[
    k_1\partial\phi_1\bar\partial\phi_1
    +k_2\partial\phi_2\bar\partial\phi_2
    +\frac{8\pi k_1k_2\lambda}{4\pi^2+k_1k_2\lambda^2}
    \partial\phi_2\,\bar\partial\phi_1
    \right].
    \label{eq:LR-two-boson-vanishing-background-action}
\end{equation}
Thus the theory at vanishing background remains a compact free-boson CFT, with
the deformation moving the theory in its constant \(G+B\) moduli.  The corresponding
metric degenerates at
\begin{equation}
    \lambda=\pm\frac{2\pi}{\sqrt{k_1k_2}},
    \label{eq:LR-two-boson-singular-locus}
\end{equation}
which is the singular locus of the flat-background finite-dimensional kernel in
this embedding.

The flat-background finite-dimensional kernel also gives the spectral form
directly.  Applying it to the undeformed product-boson trace and carrying out the
Gaussian integral over the dynamical flat holonomies gives a trace over the same
twisted Hilbert space.  Define
\begin{equation}
    \gamma_\lambda
    =
    \log\!\left(
    \frac{1+\frac{\sqrt{k_1k_2}\lambda}{2\pi}}
         {1-\frac{\sqrt{k_1k_2}\lambda}{2\pi}}
    \right).
    \label{eq:LR-two-boson-boost-parameter}
\end{equation}
For a state with undeformed twisted energy \(H_{0,\theta}\), momentum
\(P_{0,\theta}\), and selected charges \(Q^1_{L,\theta}\), \(Q^2_{R,\theta}\),
the deformation integral gives
\begin{equation}
    H_{\lambda,\theta}
    =
    H_{0,\theta}
    +
    \frac{\sinh^2\gamma_\lambda}{k_1}\bigl(Q^1_{L,\theta}\bigr)^2
    +
    \frac{\sinh^2\gamma_\lambda}{k_2}\bigl(Q^2_{R,\theta}\bigr)^2
    +
    \frac{\sinh(2\gamma_\lambda)}{\sqrt{k_1k_2}}\,
    Q^1_{L,\theta}Q^2_{R,\theta},
    \qquad
    P_{\lambda,\theta}=P_{0,\theta}.
    \label{eq:LR-two-boson-hamiltonian-flow}
\end{equation}
Thus the deformation changes the energy levels by a quadratic form in the two
selected charges, while the momentum and the conserved charges remain
those of the undeformed product theory.

\subsection{Anomaly-free massive examples}
\label{subsec:anomaly-free-massive-examples}

We next apply the anomaly-free kernel to massive theories with two vector
\(U(1)\) flavor symmetries.  The two examples below have different local
realizations: the scalar currents contain background-field, or diamagnetic,
terms, while the Dirac currents do not.  Nevertheless, in a contact-term-free
flat-background scheme their finite-volume spectra are deformed in the same
way.  In such a scheme the seed partition function is written as
\begin{equation}
    Z_0[A_{\rm flat}]
    =
    \sum_{\alpha}
    \exp\!\left[
        -2\pi\tau_2E^0_\alpha(\theta^1,\theta^2)
        +2\pi i\tau_1P^0_\alpha(\theta^1,\theta^2)
        +i\nu^a q^a_\alpha
    \right],
    \label{eq:massive-examples-contact-free-state-sum}
\end{equation}
with source-independent bulk terms absorbed into the normalization.  Applying
the \(C_{ab}=0\) kernel shifts the dynamical twists according to
\begin{equation}
    \Theta^1_\alpha=\theta^1+2\lambda q^2_\alpha,
    \qquad
    \Theta^2_\alpha=\theta^2-2\lambda q^1_\alpha,
    \label{eq:massive-examples-universal-twist-shift}
\end{equation}
and hence
\begin{equation}
    E^\lambda_\alpha(\theta^1,\theta^2)
    =
    E^0_\alpha(\Theta^1_\alpha,\Theta^2_\alpha),
    \qquad
    P^\lambda_\alpha(\theta^1,\theta^2)
    =
    P^0_\alpha(\Theta^1_\alpha,\Theta^2_\alpha),
    \qquad
    Q^a_{\lambda,\alpha}=q^a_\alpha .
    \label{eq:massive-examples-universal-spectrum}
\end{equation}
This statement is scheme-dependent in the expected sense: a flat-background
contact term is part of the input partition function and is transformed by the
same kernel.

\subsubsection{Massive complex bosons}
\label{sec:complex-boson-example}

We next apply the anomaly-free kernel to a non-conformal bosonic theory.  The
seed theory consists of two massive complex scalars with independent vector
\(U(1)\) flavor symmetries.  This example is useful because the scalar currents
depend explicitly on the background gauge fields; consequently the local
deformed Lagrangian is non-polynomial, although the flat-background partition
function is still obtained by a simple shift of twists.

\paragraph{The undeformed theory.}
\label{subsec:complex-boson-undeformed}

For \(a=1,2\), let \(\Phi_a\) be a massive complex scalar.  We couple the two
vector symmetries to background gauge fields \(A^a\), with
\begin{equation}
    D^{(a)}(A)\Phi_a=(d+iA^a)\Phi_a,
    \qquad
    D_z^{(a)}(A)\Phi_a=(\partial+iA_z^a)\Phi_a,
    \qquad
    D_{\bar z}^{(a)}(A)\Phi_a=(\bar\partial+iA_{\bar z}^a)\Phi_a .
    \label{eq:complex-boson-covariant-derivative}
\end{equation}
The undeformed action is
\begin{equation}
\begin{aligned}
    S_{\rm cb}[\Phi,\Phi^*;A^1,A^2]
    &=
    \sum_{a=1}^{2}
    \int_\Sigma d^2z\,
    \left[
        \left(D_z^{(a)}(A)\Phi_a\right)^*
        D_{\bar z}^{(a)}(A)\Phi_a
        +
        \left(D_{\bar z}^{(a)}(A)\Phi_a\right)^*
        D_z^{(a)}(A)\Phi_a
        \right.
        \\
        &\qquad\qquad\qquad
        \left.
        +
        \frac12 M_a^2|\Phi_a|^2
    \right],
    \qquad
    M_a>0 .
\end{aligned}
    \label{eq:complex-boson-seed-action}
\end{equation}
The mass terms are invariant under the vector symmetries and complex scalars
have no two-dimensional gauge anomaly.  With the gauge-transformation
convention of Section \ref{subsec:anomaly-free-transform},
\begin{equation}
    Z_{\rm cb}[A^1-\vartheta^1(U),A^2-\vartheta^2(U)]
    =
    Z_{\rm cb}[A^1,A^2].
    \label{eq:complex-boson-anomaly-free}
\end{equation}
The background-field currents are defined by
\begin{equation}
    \delta_A S_{\rm cb}
    =
    -\sum_{a=1}^{2}
    \int_\Sigma
    \delta A^a\wedge *J^a(A).
    \label{eq:complex-boson-current-definition}
\end{equation}
With the convention in \eqref{eq:complex-boson-seed-action}, this gives
\begin{equation}
    J^a(A)
    =
    i\left[
        \Phi_a^*D^{(a)}(A)\Phi_a
        -
        \left(D^{(a)}(A)\Phi_a\right)^*\Phi_a
    \right].
    \label{eq:complex-boson-current-background}
\end{equation}
Equivalently, in polar variables \(\Phi_a=R_a e^{i\varphi_a}\),
\begin{equation}
    J^a_z(A)=-2R_a^2(\partial\varphi_a+A^a_z),
    \qquad
    J^a_{\bar z}(A)=-2R_a^2(\bar\partial\varphi_a+A^a_{\bar z}) .
    \label{eq:complex-boson-current-diamagnetic}
\end{equation}
Thus, unlike the massive Dirac current below, the scalar current depends
explicitly on the background gauge field.  On shell,
\begin{equation}
    d*J^a(A)=0,
    \qquad
    a=1,2 .
    \label{eq:complex-boson-current-conservation}
\end{equation}

On the torus we use the convention of
Section~\ref{subsec:torus-flat-background-transform}.  The \(A\)-cycle
holonomies \(\theta^a\) specify the spatial twists of the two complex scalars,
while the \(B\)-cycle holonomies \(\nu^a\) are the corresponding Euclidean
chemical potentials.  Since the background coupling is anomaly-free, there is no
fixed anomaly contact term.

On a spatial circle of length \(2\pi\), the momentum lattice and one-particle
energy in the \(\theta^a\)-twisted sector are
\begin{equation}
    \kappa^a_r(\theta^a)
    =
    r+\frac{\theta^a}{2\pi},
    \qquad
    r\in\mathbb Z .
    \qquad
    \omega^a_r(\theta^a)
    =
    \sqrt{M_a^2+\left(\kappa^a_r(\theta^a)\right)^2}.
    \label{eq:complex-boson-momentum-lattice}
\end{equation}
A Fock state \(\alpha\) is specified by occupation numbers
\(N^{a,\pm}_{\alpha,r}\in\mathbb Z_{\geq0}\), where \(+\) and \(-\) denote
particles and antiparticles.  Its charge is
\begin{equation}
    q^a_\alpha
    =
    \sum_{r\in\mathbb Z}
    \left(
        N^{a,+}_{\alpha,r}
        -
        N^{a,-}_{\alpha,r}
    \right).
    \label{eq:complex-boson-charge-eigenvalue}
\end{equation}
Its undeformed energy and momentum are
\begin{equation}
\begin{aligned}
    E^0_\alpha(\theta^1,\theta^2)
    &=
    E_{\rm vac}(\theta^1,\theta^2)
    +
    \sum_{a=1}^{2}
    \sum_{r\in\mathbb Z}
    \omega^a_r(\theta^a)
    \left(
        N^{a,+}_{\alpha,r}
        +
        N^{a,-}_{\alpha,r}
    \right),
    \\
    P^0_\alpha(\theta^1,\theta^2)
    &=
    P_{\rm vac}(\theta^1,\theta^2)
    +
    \sum_{a=1}^{2}
    \sum_{r\in\mathbb Z}
    \kappa^a_r(\theta^a)
    \left(
        N^{a,+}_{\alpha,r}
        -
        N^{a,-}_{\alpha,r}
    \right).
\end{aligned}
    \label{eq:complex-boson-energy-momentum}
\end{equation}
We use the parity-symmetric scheme \(P_{\rm vac}=0\).  After subtracting the
infinite-volume bulk term, the finite-size vacuum energy is
\begin{equation}
    E_{\rm vac}(\theta^1,\theta^2)
    =
    \sum_{a=1}^{2}E_{{\rm vac},a}(\theta^a;M_a),
    \qquad
    E_{{\rm vac},a}(\theta^a;M_a)
    =
    -\frac{2M_a}{\pi}
    \sum_{w=1}^{\infty}
    \frac{\cos(w\theta^a)}{w}
    K_1(2\pi M_a w),
    \label{eq:complex-boson-vacuum-energy}
\end{equation}
where \(K_1\) is the modified Bessel function.  With these definitions, the
flat-background partition function is the state sum
\begin{equation}
\begin{aligned}
    Z_{\rm cb}
    [A^1_{\rm flat},A^2_{\rm flat}]
    =
    \sum_{\alpha\in\mathcal H_{\theta^1,\theta^2}}
    \exp\!\left[
        -2\pi\tau_2 E^0_\alpha(\theta^1,\theta^2)
        +2\pi i\tau_1 P^0_\alpha(\theta^1,\theta^2)
        +i\nu^a q^a_\alpha
    \right].
\end{aligned}
    \label{eq:complex-boson-state-sum}
\end{equation}
\paragraph{Classical deformed action.}
\label{subsec:complex-boson-classical-deformation}

We now describe the local classical deformation at vanishing background gauge
field.  In polar variables,
\begin{equation}
    \Phi_a=R_a e^{i\varphi_a},
    \qquad
    R_a\geq0,
    \qquad
    a=1,2 .
    \label{eq:complex-boson-polar-fields}
\end{equation}
The zero-background seed currents have components
\begin{equation}
    (J^a_0)_z=-2R_a^2\partial\varphi_a,
    \qquad
    (J^a_0)_{\bar z}=-2R_a^2\bar\partial\varphi_a,
    \qquad
    a=1,2 .
    \label{eq:complex-boson-zero-background-current-polar}
\end{equation}
The finite classical action at zero background is
\begin{equation}
    S_{\lambda,{\rm cb}}^{\rm cl}
    =
    \int_\Sigma d^2z\,
    \mathcal L_{\lambda,{\rm cb}}^{\rm cl},
    \label{eq:complex-boson-deformed-action-zero-background}
\end{equation}
with
\begin{equation}
\begin{aligned}
    \mathcal L_{\lambda,{\rm cb}}^{\rm cl}
    &=
    \sum_{a=1}^{2}
    \left[
        2\partial R_a\bar\partial R_a
        +
        \frac12 M_a^2R_a^2
    \right]
    \\
    &\quad+
    \frac{
        2R_1^2\partial\varphi_1\bar\partial\varphi_1
        +
        2R_2^2\partial\varphi_2\bar\partial\varphi_2
        +
        8\lambda R_1^2R_2^2
        \left(
        \partial\varphi_1\bar\partial\varphi_2
        -
        \bar\partial\varphi_1\partial\varphi_2
        \right)
    }{
        1+16\lambda^2R_1^2R_2^2
    } .
\end{aligned}
    \label{eq:complex-boson-deformed-lagrangian-polar}
\end{equation}
The background-field current of the deformed theory is defined by varying the
background gauge field before setting \(A^a=0\).  At zero background this gives
\begin{equation}
\begin{aligned}
    J^1_{\lambda}
    &=
    -\frac{2R_1^2}{1+16\lambda^2R_1^2R_2^2}
    \left[
        \left(\partial\varphi_1-4\lambda R_2^2\partial\varphi_2\right)dz
        +
        \left(\bar\partial\varphi_1+4\lambda R_2^2\bar\partial\varphi_2\right)
        d\bar z
    \right],
    \\
    J^2_{\lambda}
    &=
    -\frac{2R_2^2}{1+16\lambda^2R_1^2R_2^2}
    \left[
        \left(\partial\varphi_2+4\lambda R_1^2\partial\varphi_1\right)dz
        +
        \left(\bar\partial\varphi_2-4\lambda R_1^2\bar\partial\varphi_1\right)
        d\bar z
    \right].
\end{aligned}
    \label{eq:complex-boson-deformed-current-polar}
\end{equation}
Thus the finite deformation is a massive \(U(1)^2\)-TsT-type deformation of
the free \(\mathbb C^2\) sigma model with a quadratic potential.  At finite
\(\lambda\) the deformation is not a polynomial current-current interaction:
the angular metric is renormalized by the factor
\((1+16\lambda^2R_1^2R_2^2)^{-1}\), and a \(B\)-field-type term is generated.
At small \(\lambda\), the action expands as
\begin{equation}
\begin{aligned}
    \mathcal L_{\lambda,{\rm cb}}^{\rm cl}
    &=
    \mathcal L_{0,{\rm cb}}
    +
    8\lambda R_1^2R_2^2
    \left(
    \partial\varphi_1\bar\partial\varphi_2
    -
    \bar\partial\varphi_1\partial\varphi_2
    \right)
    +O(\lambda^2)
\end{aligned}
    \label{eq:complex-boson-small-lambda-leading}
\end{equation}
This leading term is precisely the expected double-current
deformation.

\paragraph{Kernel transform and finite-volume spectrum.}
\label{subsec:complex-boson-deformed-partition-function}

We now apply the \(C_{ab}=0\) flat-background kernel to the undeformed
partition function \eqref{eq:complex-boson-state-sum}.  For \(C_{ab}=0\), the
genus-one kernel is
\begin{equation}
    \mathcal K_{\lambda}^{(1)}
    =
    \exp\!\left[
        -\frac{i}{2\lambda}
        \left(
            \Delta\theta^1\Delta\nu^2
            -
            \Delta\nu^1\Delta\theta^2
        \right)
    \right],
    \label{eq:complex-boson-C0-kernel}
\end{equation}
where
\begin{equation}
    \Delta\theta^a=\theta^a-\widetilde\theta^a,
    \qquad
    \Delta\nu^a=\nu^a-\widetilde\nu^a .
    \label{eq:complex-boson-C0-differences}
\end{equation}
Since \(\det\mathsf M_{\lambda,0}=1/(4\lambda^2)\), the flat-background
transform becomes
\begin{equation}
\begin{aligned}
    Z_{\lambda,{\rm cb}}[A_{\rm flat}]
    =
    \frac{1}{16\pi^2\lambda^2}
    \prod_{a=1}^{2}
    \int_{-\infty}^{\infty}
    d\widetilde\theta^a\,d\widetilde\nu^a\,
    \mathcal K_{\lambda}^{(1)}
    Z_{\rm cb}[B_{\rm flat}] .
\end{aligned}
    \label{eq:complex-boson-C0-transform}
\end{equation}
Substituting the undeformed state sum, the \(\widetilde\nu^a\) integrals are
Fourier transforms.  They localize the dynamical twists to
\begin{equation}
    \Theta^1_\alpha
    =
    \theta^1+2\lambda q^2_\alpha,
    \qquad
    \Theta^2_\alpha
    =
    \theta^2-2\lambda q^1_\alpha .
    \label{eq:complex-boson-effective-twists}
\end{equation}
The Jacobian cancels the normalization in
\eqref{eq:complex-boson-C0-transform}, while the remaining part of the kernel
gives the external chemical-potential factor.  Therefore
\begin{equation}
\begin{aligned}
    Z_{\lambda,{\rm cb}}[A_{\rm flat}]
    =
    \sum_{\alpha}
    \exp\!\left[
        -2\pi\tau_2
        E^0_\alpha(\Theta^1_\alpha,\Theta^2_\alpha)
        \right.
        \left.
        +2\pi i\tau_1
        P^0_\alpha(\Theta^1_\alpha,\Theta^2_\alpha)
        +i\nu^a q^a_\alpha
    \right].
\end{aligned}
    \label{eq:complex-boson-deformed-state-sum}
\end{equation}
There is no anomaly contact term.  The spectrum is the undeformed twisted
spectrum evaluated at \(\Theta_\alpha\):
\begin{equation}
    E^\lambda_\alpha(\theta^1,\theta^2)
    =
    E^0_\alpha(\Theta^1_\alpha,\Theta^2_\alpha),
    \qquad
    P^\lambda_\alpha(\theta^1,\theta^2)
    =
    P^0_\alpha(\Theta^1_\alpha,\Theta^2_\alpha),
    \qquad
    Q^a_{\lambda,\alpha}=q^a_\alpha .
    \label{eq:complex-boson-deformed-spectrum-abstract}
\end{equation}
Thus, in a sector with fixed total charges \((q^1_\alpha,q^2_\alpha)\), the
deformed theory has the same oscillator spectrum as two free massive complex
bosons, but with charge-dependent boundary conditions
\begin{equation}
    \Phi_1(\sigma+2\pi)
    =
    e^{i(\theta^1+2\lambda q^2_\alpha)}
    \Phi_1(\sigma),
    \qquad
    \Phi_2(\sigma+2\pi)
    =
    e^{i(\theta^2-2\lambda q^1_\alpha)}
    \Phi_2(\sigma).
    \label{eq:complex-boson-charge-dependent-boundary-conditions}
\end{equation}
The local Lagrangian in Section~\ref{subsec:complex-boson-classical-deformation}
is therefore a non-polynomial massive \(U(1)^2\)-TsT-type theory whose
finite-volume spectrum is solved by the holonomy kernel.

\subsubsection{Massive Dirac fermions and the Federbush model}
\label{sec:massive-dirac-example}

The fermionic analogue is locally simpler.  The seed theory is a massive,
anomaly-free QFT with two vector flavor symmetries, and the corresponding
currents do not contain diamagnetic terms.  The finite local deformation is
therefore the Federbush-type current-current interaction, while the torus kernel
again shifts twists by the charges of the state.

\paragraph{The undeformed theory.}
\label{subsec:massive-dirac-undeformed}

Let \(\psi_a\), \(a=1,2\), be two massive Dirac fermions on a Euclidean spin
surface.  We couple the two vector symmetries to background gauge fields \(A^a\):
\begin{equation}
\begin{aligned}
    S_{\rm D}[\psi,\bar\psi;A^1,A^2]
    =
    \sum_{a=1}^{2}
    \frac12\int_{\Sigma} d^2z\,
    \bar\psi_a
    \left[
        \gamma^z
        \left(
            \nabla_z+iA^a_z
        \right)
        +
        \gamma^{\bar z}
        \left(
            \nabla_{\bar z}+iA^a_{\bar z}
        \right)
        +M_a
    \right]\psi_a ,
    \qquad
    M_a>0 .
\end{aligned}
    \label{eq:massive-dirac-seed-action}
\end{equation}
The mass terms are vector-like, so the background coupling is anomaly-free:
\begin{equation}
    Z_{\rm D}[A^1-\vartheta^1(U),A^2-\vartheta^2(U)]
    =
    Z_{\rm D}[A^1,A^2].
    \label{eq:massive-dirac-anomaly-free}
\end{equation}
The background-field currents are defined by the variation
\begin{equation}
    \delta_A S_{\rm D}
    =
    -\sum_{a=1}^{2}
    \int_\Sigma
    \delta A^a\wedge *J^a .
    \label{eq:massive-dirac-current-definition}
\end{equation}
which gives
\begin{equation}
    J^a_z=-i\,\bar\psi_a\gamma_z\psi_a,
    \qquad
    J^a_{\bar z}=-i\,\bar\psi_a\gamma_{\bar z}\psi_a .
    \label{eq:massive-dirac-current-components}
\end{equation}
The factor of \(i\) is a Euclidean convention.  The equations of motion imply
\begin{equation}
    d*J^a=0,
    \qquad
    a=1,2 .
    \label{eq:massive-dirac-current-conservation}
\end{equation}
On the torus we fix a spatial spin structure
\begin{equation}
    \alpha_{\mathfrak s}
    =
    \begin{cases}
        0, & {\rm R},\\[1mm]
        \frac12, & {\rm NS}.
    \end{cases}
    \label{eq:massive-dirac-spatial-spin}
\end{equation}
On a spatial circle of length \(2\pi\), the momentum lattice and one-particle
energy in the \(\theta^a\)-twisted sector are
\begin{equation}
    \kappa_r^a(\theta^a)
    =
    r+\frac{\theta^a}{2\pi},
    \qquad
    r\in\mathbb Z+\alpha_{\mathfrak s}.
    \label{eq:massive-dirac-momentum-lattice}
\end{equation}
\[
    \omega_r^a(\theta^a)=\sqrt{M_a^2+\left(\kappa_r^a(\theta^a)\right)^2}.
\]
A Fock state \(\alpha\) is specified by fermionic occupation numbers
\(N_{\alpha,r}^{a,\pm}\in\{0,1\}\), where \(+\) and \(-\) denote particles and
antiparticles.  Its charge is
\begin{equation}
    q^a_\alpha
    =
    \sum_{r\in\mathbb Z+\alpha_{\mathfrak s}}
    \left(
        N_{\alpha,r}^{a,+}
        -
        N_{\alpha,r}^{a,-}
    \right).
    \label{eq:massive-dirac-charge-eigenvalue}
\end{equation}
Its undeformed energy and momentum are
\begin{equation}
\begin{aligned}
    E^0_\alpha(\theta^1,\theta^2)
    &=
    E_{\rm vac}^{\mathfrak s}(\theta^1,\theta^2)
    +
    \sum_{a=1}^{2}
    \sum_{r\in\mathbb Z+\alpha_{\mathfrak s}}
    \omega_r^a(\theta^a)
    \left(
        N_{\alpha,r}^{a,+}
        +
        N_{\alpha,r}^{a,-}
    \right),
    \\
    P^0_\alpha(\theta^1,\theta^2)
    &=
    P_{\rm vac}^{\mathfrak s}(\theta^1,\theta^2)
    +
    \sum_{a=1}^{2}
    \sum_{r\in\mathbb Z+\alpha_{\mathfrak s}}
    \kappa_r^a(\theta^a)
    \left(
        N_{\alpha,r}^{a,+}
        -
        N_{\alpha,r}^{a,-}
    \right).
\end{aligned}
    \label{eq:massive-dirac-energy-momentum}
\end{equation}
We use the parity-symmetric scheme \(P_{\rm vac}^{\mathfrak s}=0\).  After
subtracting the infinite-volume bulk term, the finite-size vacuum energy is
\begin{equation}
\begin{aligned}
    E_{\rm vac}^{\mathfrak s}(\theta^1,\theta^2)
    &=
    \sum_{a=1}^{2}E_{{\rm vac},a}^{\mathfrak s}(\theta^a;M_a),
    \\
    E_{{\rm vac},a}^{\mathfrak s}(\theta^a;M_a)
    &=
    \frac{2M_a}{\pi}
    \sum_{w=1}^{\infty}
    \frac{
        \cos\!\left(2\pi w\alpha_{\mathfrak s}+w\theta^a\right)
    }{w}
    K_1(2\pi M_a w).
\end{aligned}
    \label{eq:massive-dirac-vacuum-energy}
\end{equation}
where \(K_1\) is the modified Bessel function.  The undeformed flat-background
partition function is the state sum
\begin{equation}
    Z_{\rm D}^{\mathfrak s}
    [A^1_{\rm flat},A^2_{\rm flat}]
    =
    \sum_{\alpha\in\mathcal H^{\mathfrak s}_{\theta^1,\theta^2}}
    \exp\!\left[
        -2\pi\tau_2 E^0_\alpha(\theta^1,\theta^2)
        +2\pi i\tau_1 P^0_\alpha(\theta^1,\theta^2)
        +i\nu^a q^a_\alpha
    \right].
    \label{eq:massive-dirac-state-sum}
\end{equation}

\paragraph{Classical deformed action.}
\label{subsec:massive-dirac-classical-deformation}

Because the Dirac vector currents \eqref{eq:massive-dirac-current-components}
contain no diamagnetic terms, integrating out the auxiliary gauge fields gives a
local polynomial deformation.  At vanishing background gauge fields the
Lagrangian is
\begin{equation}
\begin{aligned}
    \mathcal L_{\lambda,{\rm D}}^{\rm cl}
    &=
    \mathcal L_{0,{\rm D}}
    -
    2\lambda
    \left[
        \left(\bar\psi_1\gamma_z\psi_1\right)
        \left(\bar\psi_2\gamma_{\bar z}\psi_2\right)
        -
        \left(\bar\psi_1\gamma_{\bar z}\psi_1\right)
        \left(\bar\psi_2\gamma_z\psi_2\right)
    \right],
    \\
    \mathcal L_{0,{\rm D}}
    &=
    \frac12
    \sum_{a=1}^{2}
    \bar\psi_a
    \left(
        \gamma^z\nabla_z+\gamma^{\bar z}\nabla_{\bar z}+M_a
    \right)\psi_a .
\end{aligned}
    \label{eq:massive-dirac-zero-background-lagrangian}
\end{equation}
The deformed background-field current is
\begin{equation}
    (J^a_{\lambda})_z=-i\,\bar\psi_a\gamma_z\psi_a,
    \qquad
    (J^a_{\lambda})_{\bar z}=-i\,\bar\psi_a\gamma_{\bar z}\psi_a,
    \qquad
    a=1,2 .
    \label{eq:massive-dirac-deformed-current}
\end{equation}
This is the Federbush model \cite{Federbush:1961zz}.  In contrast with the massive complex scalar example,
the finite local action is polynomial because the Dirac vector currents contain
no diamagnetic terms.

\paragraph{Kernel transform and finite-volume spectrum.}
\label{subsec:massive-dirac-deformed-partition-function}

We now insert \eqref{eq:massive-dirac-state-sum} into the same anomaly-free
kernel transform used in \eqref{eq:complex-boson-C0-transform}.  The
\(\widetilde\nu^a\)-integrals again localize the dynamical twists to
\begin{equation}
    \Theta^1_\alpha
    =
    \theta^1+2\lambda q^2_\alpha,
    \qquad
    \Theta^2_\alpha
    =
    \theta^2-2\lambda q^1_\alpha .
    \label{eq:massive-dirac-effective-twists}
\end{equation}
The Jacobian cancels the kernel normalization, and the remaining phase gives the
external chemical-potential factor.  Thus
\begin{equation}
\begin{aligned}
    Z_{\lambda,{\rm D}}^{\mathfrak s}[A_{\rm flat}]
    =
    \sum_{\alpha}
    \exp\!\left[
        -2\pi\tau_2
        E^0_\alpha(\Theta^1_\alpha,\Theta^2_\alpha)
        \right.
        \left.
        +2\pi i\tau_1
        P^0_\alpha(\Theta^1_\alpha,\Theta^2_\alpha)
        +i\nu^a q^a_\alpha
    \right].
\end{aligned}
    \label{eq:massive-dirac-deformed-state-sum}
\end{equation}
There is no anomaly contact term.  As in the scalar example, the deformed
spectrum is the undeformed twisted spectrum evaluated at \(\Theta_\alpha\):
\begin{equation}
    E^\lambda_\alpha(\theta^1,\theta^2)
    =
    E^0_\alpha(\Theta^1_\alpha,\Theta^2_\alpha),
    \qquad
    P^\lambda_\alpha(\theta^1,\theta^2)
    =
    P^0_\alpha(\Theta^1_\alpha,\Theta^2_\alpha),
    \qquad
    Q^a_{\lambda,\alpha}=q^a_\alpha .
    \label{eq:massive-dirac-deformed-spectrum-abstract}
\end{equation}
Thus, in a sector with fixed charges, the deformed theory has the same
oscillator spectrum as two free massive Dirac fermions, but with
charge-dependent boundary conditions
\begin{equation}
    \psi_1(\sigma+2\pi)
    =
    e^{2\pi i\alpha_{\mathfrak s}}\,
    e^{i(\theta^1+2\lambda q^2_\alpha)}
    \psi_1(\sigma),
    \qquad
    \psi_2(\sigma+2\pi)
    =
    e^{2\pi i\alpha_{\mathfrak s}}\,
    e^{i(\theta^2-2\lambda q^1_\alpha)}
    \psi_2(\sigma).
    \label{eq:massive-dirac-charge-dependent-boundary-conditions}
\end{equation}
So the finite-volume spectrum is solved by the
holonomy kernel: the deformation simply replaces the external twists by the
charge-dependent twists \eqref{eq:massive-dirac-effective-twists}.

The same result can be seen directly from the excited-state TBA
\cite{Dorey:1996re}.  Similar analyses for \(T\bar T\) and \(J\bar T\)
deformations can be found in \cite{Cavaglia:2016oda,Anous:2019osb}.  Label the
one-particle species by \(A=(a,\sigma)\), with \(a=1,2\) and
\(\sigma=\pm1\).  The Federbush scattering differs from the free Dirac
scattering only by a rapidity-independent phase \cite{Castro-Alvaredo:2001lek},
\begin{equation}
    S^{\rm FB}_{(a,\sigma),(b,\tau)}
    =
    S^{\rm free}_{(a,\sigma),(b,\tau)}
    \exp\!\left[-2i\lambda\,\varepsilon_{ab}\sigma\tau\right],
    \qquad
    \varepsilon_{12}=1,
    \qquad
    \varepsilon_{21}=-1 .
    \label{eq:federbush-S-phase}
\end{equation}
Hence
\begin{equation}
    \varphi^{\rm FB}_{AB}(\vartheta)
    =
    -i\partial_\vartheta\log S^{\rm FB}_{AB}(\vartheta)
    =
    0 ,
    \label{eq:federbush-zero-kernel}
\end{equation}
so the Federbush TBA has the same free-fermion integral equation.  For an
excited state \(\alpha\), the constant source term is
\begin{equation}
    \sum_j \log S^{\rm FB}_{(a,\sigma),(a_j,\sigma_j)}
    =
    \text{fermionic branch}
    -
    2i\lambda\,\sigma\,\varepsilon_{ab}q^b_\alpha .
    \label{eq:federbush-source-term}
\end{equation}
Using this source term, the pseudoenergy can be reorganized as
\begin{equation}
\begin{aligned}
    \epsilon^{\rm FB}_{a,\sigma}(\vartheta;\theta,q_\alpha)
    &=
    \epsilon^0_{a,\sigma}(\vartheta;\theta)
    -
    2i\lambda\,\sigma\,\varepsilon_{ab}q^b_\alpha
    \\
    &=
    R M_a\cosh\vartheta
    -
    i\sigma
    \left(
        \theta^a+2\lambda\varepsilon_{ab}q^b_\alpha
    \right)
    +
    \text{state-independent branch}
    \\
    &=
    \epsilon^0_{a,\sigma}
    \bigl(\vartheta;\Theta^a_\alpha\bigr).
\end{aligned}
    \label{eq:federbush-epsilon-reorganized}
\end{equation}
Thus the Federbush pseudoenergy is the free massive Dirac pseudoenergy with
shifted twists,
\begin{equation}
    \Theta^a_\alpha
    =
    \theta^a+2\lambda\varepsilon_{ab}q^b_\alpha .
    \label{eq:federbush-effective-twist-from-TBA}
\end{equation}
Therefore
\begin{equation}
    \Theta^1_\alpha=\theta^1+2\lambda q^2_\alpha,
    \qquad
    \Theta^2_\alpha=\theta^2-2\lambda q^1_\alpha ,
    \label{eq:federbush-effective-twists-explicit}
\end{equation}
and the TBA gives
\begin{equation}
    E^\lambda_\alpha(\theta^1,\theta^2)
    =
    E^0_\alpha(\Theta^1_\alpha,\Theta^2_\alpha),
    \qquad
    P^\lambda_\alpha(\theta^1,\theta^2)
    =
    P^0_\alpha(\Theta^1_\alpha,\Theta^2_\alpha),
    \qquad
    Q^a_{\lambda,\alpha}=q^a_\alpha .
    \label{eq:federbush-spectrum-from-TBA}
\end{equation}
This reproduces the spectrum obtained in
\eqref{eq:massive-dirac-deformed-spectrum-abstract}.

\section{Non-Abelian double-current and Yang-Baxter deformations}
\label{sec:yang-baxter-generalization}

In this section we extend the path integral construction to non-Abelian
symmetries.  The deformation is not specified by the symmetry group alone: one
must choose a Lie-algebra two-cocycle
\(\omega\), or equivalently, when it is
non-degenerate on its image, the corresponding homogeneous Yang-Baxter
\(R\)-matrix.  This data replaces the constant antisymmetric tensor in the
Abelian kernel, while the relative one-form \(\Theta=A-B_h\) becomes
\(\operatorname{ad}P\)-valued.  Globally, \(\omega\) must define a fiberwise
two-form on \(\operatorname{ad}P\), which requires the corresponding
\(G_\omega\)-reduction of the background bundle.

With this data fixed, we formulate the anomaly-compatible Yang-Baxter
path integral transform.  In the absence of anomalies this is
the non-Abelian path integral construction related to Deformed T-dual Models
and homogeneous Yang-Baxter deformations
\cite{Dubovsky:2023lza,Borsato:2016pas,Borsato:2017qsx}; see also
\cite{Borsato:2018spz,Borsato:2023dis} for the relation to current-current
deformations.  In the anomalous case, the same Yang-Baxter kernel is
supplemented by parallel transport in the fixed anomaly line bundle, so that
the deformed partition function retains the finite anomaly of the seed theory.

\subsection{Non-Abelian data and global definition}
\label{subsec:yb-data-global-setup}

We first fix the background-gauge conventions.  Let \(G\) be a symmetry group
with Lie algebra \(\mathfrak g\).  We assume that the undeformed theory can be
coupled to a background \(G\)-connection, possibly with a finite anomaly.  Thus,
for a connection \(B\) and a bundle automorphism \(U\), the generating functional
obeys
\begin{equation}
    Z_0[B^U]
    =
    \exp[-\mathcal I_{\rm anom}[U,B]]\,Z_0[B],
    \label{eq:yb-undeformed-anomaly}
\end{equation}
In a local trivialization we have
\begin{equation}
    B^U
    =
    UBU^{-1}-dU\,U^{-1}.
    \label{eq:yb-gauge-transform-convention}
\end{equation}
The finite anomaly functional \(\mathcal I_{\rm anom}\) is assumed to satisfy the
Wess-Zumino consistency condition \cite{Wess:1971yu}.  The anomaly-free case is recovered by
setting \(\mathcal I_{\rm anom}=0\).

The path integral contains an external background gauge field \(A\) and a
dynamical gauge field \(B\).  As in Section \ref{subsec:anomalous-transform}, we integrate only over
the sector in which these two connections live on isomorphic bundles.  If
\(A\) and \(B\) are connections on \(P_A\to\Sigma\) and \(P_B\to\Sigma\), this
means
\begin{equation}
    [P_B]=[P_A]\in{\rm Bun}_G(\Sigma).
    \label{eq:yb-same-bundle-class}
\end{equation}
After choosing an isomorphism \(P_B\simeq P_A\), we regard both connections as
connections on a common bundle \(P\).  This bundle need not be topologically
trivial.  The Stueckelberg field is then
\begin{equation}
    h\in\Gamma(\Sigma,{\rm Ad}\,P),
    \qquad
    {\rm Ad}\,P=P\times_{\rm Ad}G,
    \label{eq:yb-stueckelberg-field}
\end{equation}
locally a \(G\)-valued function.  We define
\begin{equation}
    B_h
    =
    hBh^{-1}-dh\,h^{-1},
    \qquad
    \Theta
    =
    A-B_h .
    \label{eq:yb-Bh-Theta}
\end{equation}
Since \(A\) and \(B_h\) are connections on the same bundle, their difference is
\begin{equation}
    \Theta\in\Omega^1(\Sigma,{\rm ad}\,P)
    \label{eq:yb-Theta-adjoint-valued}
\end{equation}
and is globally defined.  If \(A\) and \(B\) were allowed to lie on
non-isomorphic bundles, the difference in \eqref{eq:yb-Bh-Theta} would not be a
globally defined adjoint-valued one-form; the deformation kernel would have to be replaced
by a pairing of relative differential-geometric objects.  We do not include
such sectors in this paper.  The auxiliary variables have the redundancy
\begin{equation}
    B\mapsto B^U,
    \qquad
    h\mapsto hU^{-1},
    \label{eq:yb-dynamical-redundancy}
\end{equation}
under which \(B_h\), and hence \(\Theta\), is invariant.

The non-Abelian deformation data include an antisymmetric two-form
\(\omega\in\wedge^2\mathfrak g^*\).  The path integral construction has the
expected current-current interpretation only when \(\omega\) is a two-cocycle,
\begin{equation}
    \omega([x,y],z)
    +
    \omega([z,x],y)
    +
    \omega([y,z],x)
    =
    0,
    \qquad
    x,y,z\in\mathfrak g .
    \label{eq:yb-cocycle-condition}
\end{equation}
Non-trivial Lie-algebra two-cocycles are equivalently non-trivial central
extensions of \(\mathfrak g\); this is the non-Abelian replacement of the
constant antisymmetric tensor in the Abelian deformation kernel.  When
\(\mathfrak g\) is equipped with an invariant
pairing \({\rm Tr}\), we can represent it by a skew-adjoint operator
\(\widetilde\omega:\mathfrak g\to\mathfrak g\),
\begin{equation}
    \omega(x,y)
    =
    \frac12{\rm Tr}\!\left[
    x\,\widetilde\omega(y)-y\,\widetilde\omega(x)
    \right],
    \label{eq:yb-omega-operator}
\end{equation}
and, when \(\widetilde\omega\) is invertible, \(R=\widetilde\omega^{-1}\)
satisfies the homogeneous classical Yang-Baxter equation
\begin{equation}
    [R(x),R(y)]
    =
    R\!\left([R(x),y]+[x,R(y)]\right).
    \label{eq:yb-cybe}
\end{equation}
Thus a nondegenerate cocycle is equivalent to the usual homogeneous
\(R\)-matrix datum \cite{Borsato:2023dis}.

There is one global condition which is invisible in the usual local presentation.
The two-cocycle must descend to a fiberwise alternating form on \({\rm ad}\,P\).
Equivalently, the transition functions of \(P\) must preserve \(\omega\), so
\(P\) must be equipped with a reduction to the stabilizer
\begin{equation}
    G_\omega
    =
    \left\{
    g\in G\,\middle|\,
    \omega({\rm Ad}_g x,{\rm Ad}_g y)=\omega(x,y)
    \ \text{for all }x,y\in\mathfrak g
    \right\}.
    \label{eq:yb-omega-stabilizer}
\end{equation}
Only under this condition is
\begin{equation}
    \omega(\Theta\wedge\Theta)
    =
    \omega(\Theta_\mu,\Theta_\nu)\,dx^\mu\wedge dx^\nu
    \label{eq:yb-omega-form-convention}
\end{equation}
a globally defined two-form on \(\Sigma\).  In the topologically trivial sector
this amounts to choosing a global trivialization together with the fixed
cocycle \(\omega\).  For a nontrivial bundle it is a restriction on the allowed
background topological class, or equivalently on the admissible background gauge
transformations.  Without this \(G_\omega\)-reduction, the Yang-Baxter deformation kernel is
not a well-defined scalar functional.

\subsection{Anomalous path integral formulation}
\label{subsec:yb-anomaly-compatible-transform}

We now define the non-Abelian double-current deformation in the presence of an
anomaly.  The topological part is the Yang-Baxter analogue of the Abelian
gauging term, while the anomalous part is the same finite parallel transport in
the anomaly line as in
Section \ref{subsec:anomalous-transform}.

Let \(\mathcal A(P)\) be the affine space of connections on the fixed bundle
\(P\).  Choose a local connection one-form \(\Omega\) on the anomaly line over
\(\mathcal A(P)\).  If \(\mathcal R_V\) denotes the finite gauge action
\(\mathcal R_V(A)=A^V\), then \(\Omega\) must transform as
\begin{equation}
    \mathcal R_V^*\Omega-\Omega
    =
    \delta_{\mathcal A}\mathcal I_{\rm anom}[V,A],
    \label{eq:yb-Omega-equivariance}
\end{equation}
where \(\delta_{\mathcal A}\) is the exterior derivative on connection space.
For a path \(\gamma_s\) in \(\mathcal A(P)\), with
\(\gamma_0=B\) and \(\gamma_1=A\), the local representative of the transport is
\begin{equation}
    T_\gamma[A,B]
    =
    \exp\!\left[
    -\int_\gamma\Omega
    \right]
    =
    \exp\!\left[
    -\int_0^1ds\,
    \Omega_{\gamma_s}(\dot\gamma_s)
    \right].
    \label{eq:yb-transport-from-connection}
\end{equation}
Assume that the path prescription is compatible with the gauge action,
\(\gamma[A^V,B^V]=\mathcal R_V\gamma[A,B]\).  Then
\eqref{eq:yb-Omega-equivariance} gives the endpoint law
\begin{equation}
    T[A^V,B^V]
    =
    \exp\!\left[
    -\mathcal I_{\rm anom}[V,A]
    +
    \mathcal I_{\rm anom}[V,B]
    \right]T[A,B],
    \qquad
    T[A,A]=1 .
    \label{eq:yb-transport-endpoint-law}
\end{equation}
This is the non-Abelian analogue of \eqref{eq:transport-endpoint-law} and
\eqref{eq:transport-from-Omega}.  From now on the path prescription is fixed and
the subscript \(\gamma\) is suppressed.  The choice of \(\Omega\) and of the path
prescription is part of the finite-\(\lambda\) definition.  Different choices
can differ by anomaly-preserving scheme factors, and the transport need not be
path independent unless the relevant anomaly-line connection has trivial
holonomy \cite{Bismut:1986wr}.

The anomaly-compatible path integral for the deformed partition function is
\begin{equation}
\begin{aligned}
    Z_{\lambda,\omega}[A]
    &=
    \int
    \frac{[D\Phi\,DB\,Dh]}
    {\mathrm{Vol}\!\left(\Gamma(\Sigma,{\rm Ad}\,P)\right)}
    \,
    T[A,B_h]
    \\
    &\quad\times
    \exp\!\left[
    -S_0[\Phi|B]
    -
    \mathcal I_{\rm anom}[h,B]
    -
    \frac{i}{4\lambda}
    \int_\Sigma\omega(\Theta\wedge\Theta)
    \right].
\end{aligned}
    \label{eq:yb-deformed-partition-function}
\end{equation}
Here \(Dh\) denotes the measure over Stueckelberg fields, i.e. sections of
\({\rm Ad}\,P\), and the quotient is by the auxiliary gauge redundancy
\eqref{eq:yb-dynamical-redundancy}.  The
factor \(\mathcal I_{\rm anom}[h,B]\) is the same finite anomaly functional as
in \eqref{eq:yb-undeformed-anomaly}, evaluated on the finite transformation
\(h\) of the connection \(B\).  In the anomaly-free case one sets
\(T=1\) and \(\mathcal I_{\rm anom}=0\). In fact, the Stueckelberg field can be gauged away, and the deformed theory is equivalent to,
\begin{equation}
    Z_{\lambda,\omega}[A]=\int [DB]T[A,B]Z_{0}[B]\exp\left[-\frac{i}{4\lambda}\omega((A-B)\wedge (A-B))\right]
\end{equation}

We first check that \eqref{eq:yb-deformed-partition-function} descends to the
auxiliary gauge quotient.  Under \eqref{eq:yb-dynamical-redundancy}, both \(B_h\) and
\(\Theta\) are invariant.  The only possible variation comes from the anomalous
matter functional and from the Stueckelberg anomaly factor.  With the convention
\((B^U)^V=B^{VU}\), the Wess-Zumino condition gives
\begin{equation}
    \mathcal I_{\rm anom}[hU^{-1},B^U]
    =
    \mathcal I_{\rm anom}[h,B]
    -
    \mathcal I_{\rm anom}[U,B].
    \label{eq:yb-dynamical-anomaly-cancellation}
\end{equation}
The anomaly produced by the transformation of \(Z_0[B]\) is therefore cancelled
by the change of \(\exp[-\mathcal I_{\rm anom}[h,B]]\).  The transport factor is
unchanged because it depends on the invariant connection \(B_h\).  Hence the
integrand is a well-defined density on the quotient by
\(\Gamma(\Sigma,{\rm Ad}\,P)\).

Next consider a background gauge transformation \(V\) which preserves the chosen
\(G_\omega\)-reduction.  Acting on the background gauge field by \(A\mapsto A^V\), we make the
change of variables
\begin{equation}
    h\mapsto Vh .
    \label{eq:yb-background-h-change}
\end{equation}
Then
\begin{equation}
    B_{Vh}=(B_h)^V,
    \qquad
    \Theta\mapsto V\Theta V^{-1}.
    \label{eq:yb-background-theta-transform}
\end{equation}
Since \(V\) preserves \(\omega\), the topological Yang-Baxter term is invariant.  The
transport endpoint law gives
\begin{equation}
    T[A^V,(B_h)^V]
    =
    \exp\!\left[
    -\mathcal I_{\rm anom}[V,A]
    +
    \mathcal I_{\rm anom}[V,B_h]
    \right]T[A,B_h],
    \label{eq:yb-background-transport}
\end{equation}
while the Wess-Zumino condition gives
\begin{equation}
    \mathcal I_{\rm anom}[Vh,B]
    =
    \mathcal I_{\rm anom}[h,B]
    +
    \mathcal I_{\rm anom}[V,B_h].
    \label{eq:yb-background-cocycle}
\end{equation}
The \(B_h\)-dependent anomaly factors cancel, leaving only the endpoint
contribution at the background gauge field:
\begin{equation}
    Z_{\lambda,\omega}[A^V]
    =
    \exp[-\mathcal I_{\rm anom}[V,A]]\,Z_{\lambda,\omega}[A].
    \label{eq:yb-deformed-anomaly}
\end{equation}
Thus the Yang-Baxter deformation preserves the finite anomaly class.  The
qualification that \(V\) preserve the \(G_\omega\)-reduction is part of the
global definition: a transformation which changes this reduction maps the theory to a
different Yang-Baxter deformation, not to another point in the same
background-gauge orbit.

\paragraph{Example: Chern-Simons anomaly.}
The preceding construction becomes explicit when the two-dimensional anomaly is
generated by a three-dimensional Chern-Simons functional \cite{Freed:1992vw}.  Let
\({\rm Tr}\) be the invariant pairing used to normalize the anomaly level, and
let \(\kappa\) obey the corresponding integrality condition.  For a
three-dimensional connection \(\widehat{\mathcal A}\), define
\begin{equation}
    {\rm CS}_3(\widehat{\mathcal A})
    =
    {\rm Tr}\!\left(
    \widehat{\mathcal A}\wedge d\widehat{\mathcal A}
    +
    \frac23
    \widehat{\mathcal A}\wedge\widehat{\mathcal A}\wedge\widehat{\mathcal A}
    \right).
    \label{eq:yb-cs-three-form}
\end{equation}
If \(Y\) is a three-manifold with \(\partial Y=\Sigma\), and if
\(\widehat{\mathcal A}\) and \(\widehat U\) extend the boundary connection
\(A\) and the boundary bundle automorphism \(U\), then a representative of the
finite anomaly is
\begin{equation}
    \mathcal I_{\rm CS}[U,A]
    =
    \frac{i\kappa}{4\pi}
    \left[
    \int_Y{\rm CS}_3(\widehat{\mathcal A}^{\widehat U})
    -
    \int_Y{\rm CS}_3(\widehat{\mathcal A})
    \right],
    \qquad
    \iota_{\partial Y}^*\widehat{\mathcal A}=A,
    \qquad
    \iota_{\partial Y}^*\widehat U=U .
    \label{eq:yb-cs-anomaly-functional}
\end{equation}
Intrinsically, \eqref{eq:yb-cs-anomaly-functional} is the ratio of two
Chern-Simons anomaly-line sections.  The displayed expression is a scalar
representative only after a local trivialization and a choice of extensions
have been made.  With
\(\widehat\vartheta(\widehat U)=d\widehat U\,\widehat U^{-1}\), the same
representative obeys
\begin{equation}
    {\rm CS}_3(\widehat{\mathcal A}^{\widehat U})
    -
    {\rm CS}_3(\widehat{\mathcal A})
    =
    d\,{\rm Tr}\!\left[
        \widehat U^{-1}d\widehat U\wedge\widehat{\mathcal A}
    \right]
    +
    \frac13{\rm Tr}\!\left[
        \left(\widehat U^{-1}d\widehat U\right)^3
    \right],
    \label{eq:yb-cs-finite-variation}
\end{equation}
which makes the Wess-Zumino consistency condition manifest as the composition
law of finite gauge transformations.

For this anomaly, a representative of the anomaly-line connection entering
\eqref{eq:yb-transport-from-connection} is the Chern-Simons symplectic
potential,
\begin{equation}
    \Omega_{{\rm CS},A}(\delta A)
    =
    -\frac{i\kappa}{4\pi}
    \int_\Sigma {\rm Tr}\!\left(A\wedge\delta A\right).
    \label{eq:yb-cs-anomaly-line-connection}
\end{equation}
Equivalently, on a closed three-manifold \(M\), equipped with a
three-dimensional connection \(\widehat{\mathcal A}\), we define
\begin{equation}
    Z_{\kappa}(M,\widehat{\mathcal A})
    =
    \exp\!\left[
    -\frac{i\kappa}{4\pi}
    \int_M {\rm CS}_3(\widehat{\mathcal A})
    \right].
    \label{eq:yb-cs-anomaly-theory}
\end{equation}
In particular, for a three-dimensional connection \(\widehat{\mathcal A}\) on
\(\Sigma\times I\) whose boundary pullbacks obey
\begin{equation}
    \iota_0^*\widehat{\mathcal A}=B,
    \qquad
    \iota_1^*\widehat{\mathcal A}=A,
    \label{eq:yb-cs-cylinder-boundary}
\end{equation}
the Chern-Simons anomaly theory gives the anomaly-line transport
\begin{equation}
    T_{{\rm CS}}[A,B;\widehat{\mathcal A}]
    =
    Z_{\kappa}(\Sigma\times I,\widehat{\mathcal A}) .
    \label{eq:yb-cs-cylinder-amplitude}
\end{equation}
Choosing temporal gauge for \(\widehat{\mathcal A}\) represents the interpolation
by a path \(\gamma_s\) of two-dimensional connections, with
\(\gamma_0=B\) and \(\gamma_1=A\).  For such a representative, this gives
\begin{equation}
    T_{{\rm CS},\gamma}[A,B]
    =
    \exp\!\left[
    \frac{i\kappa}{4\pi}
    \int_0^1ds\int_\Sigma
    {\rm Tr}\!\left(\gamma_s\wedge\dot\gamma_s\right)
    \right].
    \label{eq:yb-cs-transport}
\end{equation}
For any gauge-compatible path prescription, this transport satisfies
\begin{equation}
    T_{\rm CS}[A^V,B^V]
    =
    \exp\!\left[
    -\mathcal I_{\rm CS}[V,A]
    +
    \mathcal I_{\rm CS}[V,B]
    \right]T_{\rm CS}[A,B].
    \label{eq:yb-cs-transport-law}
\end{equation}
A useful local scheme is the straight path
\begin{equation}
    \gamma_s=B+s(A-B).
    \label{eq:yb-cs-straight-path}
\end{equation}
Since \(A\) and \(B\) are connections on the same bundle, the difference
\(A-B\in\Omega^1(\Sigma,{\rm ad}\,P)\) is global, and
\eqref{eq:yb-cs-straight-path} is well-defined.  In the same local
trivialization as above, this gives the representative
\begin{equation}
    T_{\rm CS}^{\rm str}[A,B]
    =
    \exp\!\left[
    \frac{i\kappa}{4\pi}
    \int_\Sigma{\rm Tr}\!\left(B\wedge(A-B)\right)
    \right].
    \label{eq:yb-cs-straight-transport}
\end{equation}
Substituting this into
\eqref{eq:yb-deformed-partition-function} gives
\begin{equation}
\begin{aligned}
    Z_{\lambda,\omega}^{\rm CS}[A]
    &=
    \int
    \frac{[D\Phi\,DB\,Dh]}
    {\mathrm{Vol}\!\left(\Gamma(\Sigma,{\rm Ad}\,P)\right)}
    \\
    &\quad\times
    \exp\!\left[
    -S_0[\Phi|B]
    -
    \mathcal I_{\rm CS}[h,B]
    +
    \frac{i\kappa}{4\pi}
    \int_\Sigma{\rm Tr}\!\left(B_h\wedge\Theta\right)
    -
    \frac{i}{4\lambda}
    \int_\Sigma\omega(\Theta\wedge\Theta)
    \right].
\end{aligned}
    \label{eq:yb-cs-deformed-partition-function}
\end{equation}
Equations \eqref{eq:yb-cs-transport-law} and the Wess-Zumino condition imply
that, for background gauge transformations preserving the chosen
\(G_\omega\)-reduction,
\begin{equation}
    Z_{\lambda,\omega}^{\rm CS}[A^V]
    =
    \exp[-\mathcal I_{\rm CS}[V,A]]\,
    Z_{\lambda,\omega}^{\rm CS}[A].
    \label{eq:yb-cs-deformed-anomaly}
\end{equation}
Thus the Chern-Simons anomaly gives an explicit realization of
\eqref{eq:yb-deformed-partition-function} without assuming that \(P\) is
topologically trivial.  The only additional restriction is the one already
stated in Section \ref{subsec:yb-data-global-setup}: the two-cocycle
\(\omega\) must define a fiberwise pairing on \({\rm ad}\,P\).

\subsection{Classical current-current flow}
\label{subsec:classical-yang-baxter-flow}

We now spell out the sense in which
\eqref{eq:yb-deformed-partition-function} defines a Yang-Baxter
current-current deformation.  The statement is classical: the dynamical gauge
field is evaluated at its saddle, and the \(\lambda\)-flow is extracted from the
explicit \(\lambda\)-dependence of the quadratic term.  The anomaly
data enter through the transport factor and therefore modify the relation
between the current appearing in this topological term and the current obtained
by differentiating with respect to the background gauge field.

Work in a local patch of field space and fix the unitary gauge \(h=1\).
Then \(B_h=B\), \(\Theta=A-B\), and the saddle action is
\begin{equation}
    S_{\rm cl}
    =
    S_0[\Phi|B]
    +
    S_T[A,B]
    +
    \frac{i}{4\lambda}
    \int_\Sigma\omega\!\left((A-B)\wedge(A-B)\right),
    \qquad
    S_T[A,B]\equiv-\log T[A,B].
    \label{eq:yb-classical-action}
\end{equation}
The matter current and the endpoint variation of the anomaly transport are
defined by
\begin{equation}
\begin{aligned}
    \delta_BS_0[\Phi|B]
    &=
    -\int_\Sigma{\rm Tr}\!\left(\delta B\wedge *\mathscr J(B)\right),
    \\
    \delta S_T[A,B]
    &=
    \int_\Sigma{\rm Tr}\!\left(\delta A\wedge\Pi^A[A,B]\right)
    +
    \int_\Sigma{\rm Tr}\!\left(\delta B\wedge\Pi^B[A,B]\right).
\end{aligned}
    \label{eq:yb-current-and-transport-response}
\end{equation}
Using the operator \(\widetilde\omega\) in \eqref{eq:yb-omega-operator}, define
the current associated with the quadratic term by
\begin{equation}
    \mathscr K_{\lambda,\omega}(A)
    \equiv
    -
    \frac{i}{2\lambda}\,
    \widetilde\omega(A-B_*),
    \label{eq:yb-kernel-current}
\end{equation}
where \(B_*\) denotes the auxiliary saddle.  The \(B\)-equation of motion gives
\begin{equation}
    \mathscr K_{\lambda,\omega}(A)
    =
    *\mathscr J(B_*)
    -
    \Pi^B[A,B_*].
    \label{eq:yb-saddle-current-relation}
\end{equation}
Let \(S_{\lambda,\omega}[\Phi|A]\) be the on-shell action obtained by solving for
the auxiliary saddle \(B_*\) in \eqref{eq:yb-classical-action}.  The physical
current of the deformed theory is defined by its response to the background
gauge field,
\begin{equation}
    \delta_AS_{\lambda,\omega}[\Phi|A]
    =
    -
    \int_\Sigma{\rm Tr}\!\left(
    \delta A\wedge *\mathscr J_{\lambda,\omega}(A)
    \right),
    \qquad
    *\mathscr J_{\lambda,\omega}(A)
    =
    \mathscr K_{\lambda,\omega}(A)-\Pi^A[A,B_*].
    \label{eq:yb-physical-current}
\end{equation}
Thus the current which appears in the Yang-Baxter flow equation is
\(\mathscr K_{\lambda,\omega}\).  The physical background current differs from it by
the endpoint response of the anomaly transport.  In an anomaly-free scheme, or
in a scheme where this endpoint response vanishes at the background of interest,
the two currents coincide.

Because \(B_*\) obeys the saddle equation, its implicit \(\lambda\)-dependence
does not contribute to the derivative of the on-shell action.  Hence
\begin{equation}
    \frac{d}{d\lambda}S_{\lambda,\omega}[\Phi|A]
    =
    -
    \frac{i}{4\lambda^2}
    \int_\Sigma\omega\!\left((A-B_*)\wedge(A-B_*)\right).
    \label{eq:yb-lambda-flow-before-R}
\end{equation}
This already gives the flow in cocycle form.  If \(\widetilde\omega\) is
invertible, \(R=\widetilde\omega^{-1}\), and \eqref{eq:yb-kernel-current}
rewrites it as
\begin{equation}
    \frac{d}{d\lambda}S_{\lambda,\omega}[\Phi|A]
    =
    i\int_\Sigma{\rm Tr}\!\left[
    R\!\left(\mathscr K_{\lambda,\omega}\right)
    \wedge
    \mathscr K_{\lambda,\omega}
    \right].
    \label{eq:yb-classical-current-current-flow}
\end{equation}
This is the homogeneous Yang-Baxter analogue of the Abelian current-current
flow in \eqref{eq:anomalous-classical-current-current-flow}.  

\section{Conclusions and outlook}
\label{sec:conclusions}

We have formulated double-current deformations of two-dimensional QFTs whose
background-field partition functions are anomalous.  The basic point is that,
in an anomalous theory, the partition function is not an ordinary function on
the space of background gauge fields, but a local representative of a section
of an anomaly line bundle.  The path integral definition of the deformation
must therefore be performed with this line bundle held fixed.  Concretely, the
usual  double-current kernel is supplemented by a Stueckelberg anomaly
factor and by parallel transport in the anomaly line bundle between the fibers
over the dynamical and background gauge fields.  The endpoint law of this
transport is fixed by the finite anomaly cocycle, and ensures that the deformed
partition function transforms with the same background anomaly as the seed
theory.

For Abelian deformations we gave an explicit path integral construction.  The
compact Stueckelberg fields make both finite and large auxiliary gauge
transformations manifest.  In the same-class sector, where the background and
dynamical gauge fields are connections on bundles with the same Chern class,
the relative field \(\Theta\) is an ordinary globally defined one-form, and the
quadratic topological kernel is well defined as a differential-form integral.
This is the sector in which the deformation parameter can be treated as a
continuous coupling.  If one instead includes dynamical gauge fields in
inequivalent bundle classes, the topological kernel should be replaced by its
differential-cohomological refinement.

For flat background gauge fields, the path integral transform reduces to an
explicit holonomy integral.  On the torus, the non-zero modes of the compact
Stueckelberg fields impose the flatness of the dynamical gauge field; after
local determinants are absorbed into the normalization, one is left with a
finite-dimensional kernel acting on real lifts of the flat holonomies.  The
same local argument applies on a genus-\(g\) Riemann surface, where the
genus-one symplectic pairing is replaced by the canonical Darboux pairing on
harmonic one-forms.  The remaining background-independent factors are
topological normalization data.  We fixed them by requiring that the
flat-background transform reduce to the identity in the undeformed limit.

The examples in Section 4 provide concrete checks of the formalism.  For the
compact boson, the deformation gives the ordinary compact-boson spectrum with
the radius parameter shifted from \(k\) to \(K_\lambda\) at vanishing background
gauge fields.  With flat background holonomies, however, the answer is not
obtained by replacing \(k\) by \(K_\lambda\) everywhere: in the fixed anomaly
scheme used here, the energy levels are deformed through \(K_\lambda\), while
the contact terms and spectral-flow charge assignments remain governed by the
original anomaly coefficient \(k\).  We also considered anomaly-free massive
examples.  For massive complex bosons and massive Dirac fermions, the
flat-background kernel shifts the twists by the charges of the state; in the
Dirac case this gives the Federbush model and matches its rapidity-independent
scattering phase.

We also formulated the non-Abelian extension underlying homogeneous
Yang-Baxter deformations.  The additional input is a Lie-algebra two-cocycle
\(\omega\), equivalently a homogeneous Yang-Baxter \(R\)-matrix on its image.
Globally, this data defines the topological kernel only after reducing the
adjoint bundle to the subgroup that preserves \(\omega\).  Once this
\(G_\omega\)-reduction is fixed, the anomaly-compatible construction proceeds
as in the Abelian case: the Yang-Baxter kernel is supplemented by parallel
transport in the fixed anomaly line bundle, with endpoint law determined by the
finite anomaly cocycle.  The resulting partition function therefore carries
the same background anomaly as the seed theory, while the saddle-point
\(\lambda\)-flow is generated by the Yang-Baxter bilinear of the on-shell
kernel current.

There are several future directions
\begin{itemize}
\item
A fully global formulation should replace the differential-form expression for
the topological kernel by its differential-cohomological refinement.  This
would allow the background and dynamical gauge fields to lie in different
bundle classes, include all topological sectors, and incorporate possible
global anomalies.

\item
It would be useful to study explicit non-Abelian examples, in particular WZW
models beyond Cartan or Abelian current-current deformations.  Such examples
should clarify how the anomaly-line-bundle construction acts on affine
representation data and on the corresponding current-algebra characters.

\item
The extension to correlation functions, boundaries, defects, and supersymmetric
backgrounds should reveal further consequences of the distinction between the
kernel current and the physical background-field current.  In these settings
the endpoint response of the anomaly-line-bundle transport is expected to
modify Ward identities, charge assignments, or boundary and defect gluing
conditions.
\end{itemize}

    \section*{Acknowledgements}
We thank Olalla A. Castro-Alvaredo, Liangyu Chen, Ho Tat Lam, Kangning Liu and Wei Song for useful discussions. 
The work is supported by the NSFC special fund for theoretical physics No. 12447108 and the national key research and development program of China No. 2020YFA0713000.

\appendix

\section{Flat-background path integral details}
\label{app:flat-background-path integral-details}

In this appendix, we give some details in the derivation of  the flat-background path integral transforms in
Section \ref{sec:partition-functions}.

\subsection{Details of the torus path integral}
\label{subsec:app-torus-path integral}

This subsection records the gauge fixing behind the flat-background
path integral transform
\eqref{eq:torus-flat-kernel-transform}.  We work in the same topological class as
the flat background gauge field.  Since a flat \(U(1)\) background gauge field
has vanishing Chern class, we can choose a trivialization in which the dynamical
gauge fields are ordinary globally defined one-forms.  The matter fields have
already been integrated out, and the starting point is
\eqref{eq:torus-dynamical-path integral}.  For brevity, write the auxiliary
exponent as
\begin{equation}
    \mathcal E[B,h;A_{\rm flat}]
    =
    \mathcal W_C[h,B]
    -\frac{i}{4\lambda}\int_{T^2}\epsilon_{ab}\Theta^a\wedge\Theta^b
    +C_{ab}\int_{T^2}A_{\rm flat}^a\wedge\Theta^b
    -\frac12 C_{ab}\int_{T^2}\Theta^a\wedge\Theta^b .
    \label{eq:app-torus-auxiliary-exponent}
\end{equation}
Here \(\Theta^a=A_{\rm flat}^a-B^a+\vartheta(h^a)\), as in
\eqref{eq:torus-Theta}.

We first fix the small gauge transformations.  With
\(\vartheta(U_0^a)=d\varrho^a\), insert
\begin{equation}
    1
    =
    \det\nolimits'\Delta_0
    \int D'\varrho^a\,
    \delta'\!\left[d^\dagger(B^a-d\varrho^a)\right],
    \qquad
    \Delta_0=d^\dagger d .
    \label{eq:app-torus-fp-identity}
\end{equation}
The prime means that the constant scalar zero mode is omitted.  In the same
class as the flat background gauge field, the Hodge decomposition is
\begin{equation}
    B^a=B^a_{\rm harm}+d\chi^a+*d\psi^a,
    \qquad
    a=1,2,
    \label{eq:app-torus-hodge-decomposition}
\end{equation}
with \(\chi^a\) and \(\psi^a\) single-valued and without constant modes.  The
gauge condition is
\begin{equation}
    d^\dagger B^a=\Delta_0\chi^a,
    \label{eq:app-torus-gauge-condition}
\end{equation}
and hence sets \(\chi^a=0\).  The one-form measure and the delta functional give
\begin{equation}
    DB^a=
    (\det\nolimits'\Delta_0)\,
    DB_{\rm harm}^a\,D'\chi^a\,D'\psi^a,
    \qquad
    \delta'[d^\dagger B^a]
    =
    (\det\nolimits'\Delta_0)^{-1}\delta'[\chi^a].
    \label{eq:app-torus-hodge-jacobian}
\end{equation}
Together with \eqref{eq:app-torus-fp-identity}, this leaves one
\(\det'\Delta_0\) for each gauge field.  After the non-constant small gauge
volume has been cancelled, the path integral becomes
\begin{equation}
\begin{aligned}
    Z_{\lambda,C}[A_{\rm flat}]
    =
    \frac{(\det'\Delta_0)^2}
    {\mathrm{Vol}(U(1)_0)^2\,\mathrm{Vol}(\mathcal G_{\rm lg})}
    \int\prod_{a=1}^{2}
    [DB_{\rm harm}^a\,D'\psi^a\,Dh^a]\,
    Z_0[B]\,
    e^{\mathcal E[B,h;A_{\rm flat}]},
\end{aligned}
     \label{eq:app-torus-after-small-gauge-fixing}
\end{equation}
where now \(B^a=B^a_{\rm harm}+*d\psi^a\).

The compact Stueckelberg field decomposes as
\begin{equation}
    \vartheta(h^a)=dX^a+\eta_{\rm w}^a,
    \qquad
    \eta_{\rm w}^a=2\pi(r^a\rho+s^a\sigma),
    \qquad
    r^a,s^a\in\mathbb Z .
    \label{eq:app-torus-h-winding}
\end{equation}
The pure large gauge transformations set \(\eta_{\rm w}^a=0\).  Their action on
the harmonic part of \(B^a\) only shifts the lifted holonomies by integral
periods; since the \(B\)-integration is over the full affine space of real
connections, this is a change of variables.  No continuous Faddeev-Popov
determinant is produced.  The remaining fields are therefore
\begin{equation}
    B^a=B^a_{\rm harm}+*d\psi^a,
    \qquad
    \vartheta(h^a)=dX^a ,
    \label{eq:app-torus-gauge-fixed-fields}
\end{equation}
and \eqref{eq:app-torus-after-small-gauge-fixing} no longer contains
\(\mathrm{Vol}(\mathcal G_{\rm lg})\).

For $\mathcal W_C[h,B]$ we have:
\begin{equation}
    \mathcal W_C[h,B]
    =
    -C_{ab}\int_{T^2}dX^a\wedge B^b .
    \label{eq:app-torus-WC-boundary}
\end{equation}
Using \(\Theta^a=A_{\rm flat}^a-B^a+dX^a\), the \(X^a\)-dependent part of the
exponent is
\begin{equation}
    \mathcal E_X
    =
    \int_{T^2} dX^c\wedge
    \left[
    \frac{i}{2\lambda}\epsilon_{ac}(A_{\rm flat}^a-B^a)
    -
    \frac12(C_{ac}+C_{ca})(A_{\rm flat}^a+B^a)
    \right].
    \label{eq:app-torus-X-linear-coupling}
\end{equation}
The compact constant mode of \(X^a\) decouples.  The non-zero modes impose
\begin{equation}
    d\left[
    \frac{i}{2\lambda}\epsilon_{ac}(A_{\rm flat}^a-B^a)
    -
    \frac12(C_{ac}+C_{ca})(A_{\rm flat}^a+B^a)
    \right]=0,
    \qquad
    c=1,2 .
    \label{eq:app-torus-X-constraint}
\end{equation}
For flat \(A_{\rm flat}^a\), this is precisely
\eqref{eq:torus-flatness-matrix}.  Thus, when
\(\det\mathsf M_{\lambda,C}\neq0\), the dynamical gauge field is localized to
\begin{equation}
    dB^1=dB^2=0 .
    \label{eq:app-torus-flatness}
\end{equation}
Since \(dB^a=d*d\psi^a=-(\Delta_0\psi^a)\,{\rm vol}_{T^2}\), flatness sets
\(\psi^a=0\) on the non-zero-mode subspace.

It remains to keep the determinant of this Fourier transform.  Let
\(\Delta_0 f_n=\lambda_n f_n\), \(\lambda_n>0\), and expand
\begin{equation}
    X_{\rm nz}^a=\sum_{n\ne0}X_n^a f_n,
    \qquad
    \psi^a=\sum_{n\ne0}\psi_n^a f_n .
    \label{eq:app-torus-mode-expansion}
\end{equation}
The non-zero-mode coupling takes the form
\begin{equation}
    \mathcal E_X
    =
    i\sum_{n\ne0}\lambda_n
    (X_n^1,X_n^2)\,
    \mathsf M_{\lambda,C}
    \begin{pmatrix}
        \psi_n^1 \\ \psi_n^2
    \end{pmatrix}.
    \label{eq:app-torus-mode-coupling}
\end{equation}
For each mode,
\begin{equation}
    \int dX_n^1\,dX_n^2\,
    e^{\,i\lambda_n X_n^{\mathsf T}\mathsf M_{\lambda,C}\psi_n}
    =
    \frac{(2\pi)^2}
    {\lambda_n^2|\det\mathsf M_{\lambda,C}|}\,
    \delta(\psi_n^1)\delta(\psi_n^2).
    \label{eq:app-torus-single-mode-integral}
\end{equation}
The product of the \(\lambda_n^{-2}\) factors gives
\((\det'\Delta_0)^{-2}\).  The mode-independent product is regularized on the
torus by the regularized number of non-zero scalar modes, \(-1\):
\begin{equation}
    \prod_{n\ne0}
    \frac{(2\pi)^2}{|\det\mathsf M_{\lambda,C}|}
    =
    \frac{|\det\mathsf M_{\lambda,C}|}{(2\pi)^2}.
    \label{eq:app-torus-mode-independent-factor}
\end{equation}
The compact constant modes of \(X^1,X^2\) cancel
\(\mathrm{Vol}(U(1)_0)^2\).  Therefore
\begin{equation}
    \frac{1}{\mathrm{Vol}(U(1)_0)^2}
    \int[DX^1DX^2]\,e^{\mathcal E_X}
    =
    \frac{|\det\mathsf M_{\lambda,C}|}{(2\pi)^2}
    (\det\nolimits'\Delta_0)^{-2}
    \delta'[\psi^1]\delta'[\psi^2].
    \label{eq:app-torus-X-functional-integral}
\end{equation}
This cancels the two scalar Laplacian determinants in
\eqref{eq:app-torus-after-small-gauge-fixing}.  Hence no additional propagating
degree of freedom is produced by the auxiliary sector.

The remaining connection is flat and equals its harmonic part,
\begin{equation}
    B_{\rm flat}^a=B_{\rm harm}^a
    =
    \widetilde\theta^a\rho+\widetilde\nu^a\sigma,
    \qquad
    \widetilde\theta^a,\widetilde\nu^a\in\mathbb R .
    \label{eq:app-torus-flat-b}
\end{equation}
For harmonic one-forms
\(\alpha=\theta_\alpha\rho+\nu_\alpha\sigma\) and
\(\beta=\theta_\beta\rho+\nu_\beta\sigma\),
\begin{equation}
    \int_{T^2}\alpha\wedge\beta
    =
    \theta_\alpha\nu_\beta-\nu_\alpha\theta_\beta .
    \label{eq:app-torus-harmonic-pairing}
\end{equation}
Substituting
\(A_{\rm flat}^a-B_{\rm flat}^a
=\Delta\theta^a\rho+\Delta\nu^a\sigma\) into the quadratic topological term and
the anomaly-line transport factor gives exactly \eqref{eq:torus-flat-kernel}.  With the
determinant factor in \eqref{eq:app-torus-X-functional-integral}, one obtains
\eqref{eq:torus-flat-kernel-transform}.

\subsection{Higher-genus derivation}
\label{subsec:app-higher-genus-derivation}

We now repeat the same gauge fixing on a closed Riemann surface \(\Sigma_g\).
The background gauge fields are flat and in the same bundle class as the
dynamical gauge fields.  The small-gauge fixing is local and is therefore
identical to Appendix \ref{subsec:app-torus-path integral}: after imposing
\(d^\dagger B^a=0\), one can write
\begin{equation}
    B^a=B_{\rm harm}^a+*d\psi^a,
    \qquad
    a=1,2,
    \label{eq:app-higher-genus-small-gauge-fixed-b}
\end{equation}
and the gauge-fixed measure contains the factor
\begin{equation}
    \bigl(\det\nolimits'\Delta_0\bigr)^2 .
    \label{eq:app-higher-genus-small-gauge-determinant}
\end{equation}
Here \(\Delta_0=d^\dagger d\) acts on scalar non-zero modes on
\(\Sigma_g\).

The compact Stueckelberg field decomposes into a single-valued lift and harmonic
winding,
\begin{equation}
    \vartheta(h^a)
    =
    dX^a
    +
    2\pi\sum_{I=1}^{g}
    (r_I^a\rho_I+s_I^a\sigma_I),
    \qquad
    r_I^a,s_I^a\in\mathbb Z .
    \label{eq:app-higher-genus-h-winding}
\end{equation}
The pure large gauge transformations set the integer winding to zero.  The
simultaneous shift of \(B_{\rm harm}^a\) only changes the lifted harmonic
coordinates, so the remaining harmonic integral is over
\(\widetilde\theta_I^a,\widetilde\nu_I^a\in\mathbb R\), not over a fundamental
domain.  After small and large gauge fixing one has
\begin{equation}
\begin{aligned}
    Z_{\lambda,C}^{(g)}[A_{\rm flat}]
    =
    \frac{(\det'\Delta_0)^2}{\mathrm{Vol}(U(1)_0)^2}
    \int\prod_{a=1}^{2}
    [DB_{\rm harm}^a\,D'\psi^a\,DX^a]\,
    Z_0^{(g)}[B]\,
    e^{\mathcal E[B,h;A_{\rm flat}]},
\end{aligned}
    \label{eq:app-higher-genus-after-gauge-fixing}
\end{equation}
where \(B^a=B_{\rm harm}^a+*d\psi^a\),
\(\vartheta(h^a)=dX^a\), and \(\mathcal E\) is the same auxiliary exponent as in
\eqref{eq:app-torus-auxiliary-exponent}, with \(T^2\)
replaced by \(\Sigma_g\).

For $\mathcal W_C[h,B]$ we have
\begin{equation}
    \mathcal W_C[h,B]
    =
    -C_{ab}\int_{\Sigma_g}dX^a\wedge B^b .
    \label{eq:app-higher-genus-WC-boundary}
\end{equation}
The \(X\)-dependent part of the exponent is therefore
\begin{equation}
    \mathcal E_X
    =
    \int_{\Sigma_g}dX^c\wedge
    \left[
    \frac{i}{2\lambda}\epsilon_{ac}(A_{\rm flat}^a-B^a)
    -
    \frac12(C_{ac}+C_{ca})(A_{\rm flat}^a+B^a)
    \right].
    \label{eq:app-higher-genus-X-linear-coupling}
\end{equation}
The compact constant modes of \(X^a\) decouple.  The non-zero modes impose
\begin{equation}
    d\left[
    \frac{i}{2\lambda}\epsilon_{ac}(A_{\rm flat}^a-B^a)
    -
    \frac12(C_{ac}+C_{ca})(A_{\rm flat}^a+B^a)
    \right]=0,
    \qquad
    c=1,2 .
    \label{eq:app-higher-genus-X-constraint}
\end{equation}
For flat background gauge fields this is
\begin{equation}
    \mathsf M_{\lambda,C}
    \begin{pmatrix}
        dB^1 \\ dB^2
    \end{pmatrix}
    =
    0,
    \label{eq:app-higher-genus-flatness-matrix}
\end{equation}
with the same matrix \(\mathsf M_{\lambda,C}\) as in
\eqref{eq:torus-flatness-matrix}.  When
\(\det\mathsf M_{\lambda,C}\neq0\), this localizes the dynamical gauge field to
the flat slice
\begin{equation}
    dB^1=dB^2=0,
    \qquad
    \psi^1=\psi^2=0 .
    \label{eq:app-higher-genus-flatness}
\end{equation}

The non-zero-mode determinant is also the same local Fourier transform as on the
torus.  If \(\Delta_0 f_n=\lambda_n f_n\), \(\lambda_n>0\), then
\begin{equation}
    \mathcal E_X
    =
    i\sum_{n\ne0}\lambda_n
    (X_n^1,X_n^2)\mathsf M_{\lambda,C}
    \begin{pmatrix}
        \psi_n^1 \\ \psi_n^2
    \end{pmatrix},
    \label{eq:app-higher-genus-mode-coupling}
\end{equation}
and each mode gives
\begin{equation}
    \int dX_n^1\,dX_n^2\,
    e^{\,i\lambda_n X_n^{\mathsf T}\mathsf M_{\lambda,C}\psi_n}
    =
    \frac{(2\pi)^2}
    {\lambda_n^2|\det\mathsf M_{\lambda,C}|}\,
    \delta(\psi_n^1)\delta(\psi_n^2).
    \label{eq:app-higher-genus-single-mode-integral}
\end{equation}
Thus the \(\lambda_n^{-2}\) product cancels the factor
\((\det'\Delta_0)^2\) in \eqref{eq:app-higher-genus-after-gauge-fixing}, while
the constant modes of \(X^1,X^2\) cancel \(\mathrm{Vol}(U(1)_0)^2\).  The only
remaining non-zero-mode factor is background independent \cite{OSGOOD1988148,Vassilevich:2003xt},
\begin{equation}
    \mathcal D_g(\lambda,C)
    =
    \left[
    \frac{(2\pi)^2}{|\det\mathsf M_{\lambda,C}|}
    \right]^{\zeta_{\Delta_0}^{(g)}(0)} ,
    \qquad
    \zeta_{\Delta_0}^{(g)}(0)
    =
    \frac{\chi(\Sigma_g)}{6}-1 .
    \label{eq:app-higher-genus-background-independent-determinant}
\end{equation}
This factor is part of the background-independent measure convention.  More
generally, the definition of the path integral transform can be multiplied by a
local topological factor depending only on the worldsheet topology,
\begin{equation}
    \exp\!\left[
        -\frac{\mu_\lambda}{4\pi}
        \int_{\Sigma_g} R_\Sigma\,\mathrm{vol}_\Sigma
    \right]
    =
    \exp\!\left[-\mu_\lambda\,\chi(\Sigma_g)\right],
    \label{eq:app-higher-genus-topological-normalization}
\end{equation}
where \(\mu_\lambda\) is a theory-dependent and scheme-dependent number.  The
Euler factor displayed above is independent of the background gauge fields, and
therefore cannot modify the finite gauge anomaly.  It is a purely topological
normalization of the path integral and is not fixed by the deformation itself.
In the main text we choose this topological normalization,
together with the harmonic measure, by the
undeformed limit of the flat-background transform.  This gives the factor
\([|\det\mathsf M_{\lambda,C}|/(2\pi)^2]^g\) in
\eqref{eq:higher-genus-flat-kernel-transform}.

It remains to evaluate the harmonic part of the exponent.  On the localized
slice,
\begin{equation}
    B_{\rm flat}^a=B_{\rm harm}^a
    =
    \sum_{I=1}^g
    (\widetilde\theta_I^a\rho_I+\widetilde\nu_I^a\sigma_I).
    \label{eq:app-higher-genus-flat-b}
\end{equation}
Using the pairing \eqref{eq:higher-genus-harmonic-pairing}, the
quadratic topological term and anomaly-line transport factor reduce to
\eqref{eq:higher-genus-flat-kernel}.  Combining this harmonic phase with the
normalization fixed above gives
\eqref{eq:higher-genus-flat-kernel-transform}.

\bibliographystyle{JHEP}
\bibliography{ref.bib}
\end{document}

%% file: preamble.tex
\usepackage{jheppub}
\usepackage{mathtools}
\usepackage{float}
\usepackage{tikz-cd}

\date{}

 \usepackage[normalem]{ulem}
\usepackage[textsize=scriptsize,textwidth=2.5cm]{todonotes}

\let\exporig\exp
\usepackage{physics}
\usepackage{enumitem}
\let\exp\exporig

\affiliation[\ensuremath{\gamma}]{Yau Mathematical Sciences Center, Tsinghua University, Beijing 100084, China}
\affiliation[\ensuremath{\tau}]{Department of Mathematical Sciences, Tsinghua University, Beijing 100084, China}

\usepackage{xspace}